\documentclass[prc,twocolumn,showpacs,floatfix,nofootinbib,preprintnumbers,superscriptaddress,amsmath,amssymb]{revtex4}

\usepackage{amsmath}           
\usepackage{amsfonts}
\usepackage{graphicx}          
\usepackage{dcolumn}           
\usepackage{bm}

\usepackage{color}

\begin{document}

\bibliographystyle{apsrev}


\title{Nuclear Energy Density Optimization}

\author{M. Kortelainen}
\affiliation{Department of Physics and Astronomy, University of Tennessee, Knoxville, TN 37996, USA}
\affiliation{Physics Division, Oak Ridge National Laboratory, P.O. Box 2008, Oak Ridge, TN 37831, USA}

\author{T. Lesinski}
\affiliation{Department of Physics and Astronomy, University of Tennessee, Knoxville, TN 37996, USA}
\affiliation{Physics Division, Oak Ridge National Laboratory, P.O. Box 2008, Oak Ridge, TN 37831, USA}

\author{J. Mor\'e}
\affiliation{Mathematics and Computer Science Division, Argonne National Laboratory, Argonne, IL 60439, USA}

\author{W. Nazarewicz}
\affiliation{Department of Physics and Astronomy, University of Tennessee, Knoxville, TN 37996, USA}
\affiliation{Physics Division, Oak Ridge National Laboratory, P.O. Box 2008, Oak Ridge, TN 37831, USA}
\affiliation{Institute of Theoretical Physics, Warsaw University, ul. Ho\.{z}a 69, PL-00681, Warsaw, Poland}

\author{J. Sarich}
\affiliation{Mathematics and Computer Science Division, Argonne National Laboratory, Argonne, IL 60439, USA}

\author{N. Schunck}
\affiliation{Department of Physics and Astronomy, University of Tennessee, Knoxville, TN 37996, USA}
\affiliation{Physics Division, Oak Ridge National Laboratory, P.O. Box 2008, Oak Ridge, TN 37831, USA}

\author{M. V.~Stoitsov}
\affiliation{Department of Physics and Astronomy, University of Tennessee, Knoxville, TN 37996, USA}
\affiliation{Physics Division, Oak Ridge National Laboratory, P.O. Box 2008, Oak Ridge, TN 37831, USA}
\affiliation{Institute of Nuclear Research and Nuclear Energy, Bulgarian Academy of Sciences, Sofia, Bulgaria}

\author{S. Wild}
\affiliation{Mathematics and Computer Science Division, Argonne National Laboratory, Argonne, IL 60439, USA}

\date{\today}

\newcommand{\cS}{\mathcal{S}}	
\newcommand{\cN} {\mbox{$\cal N$}} 
\newcommand{\Ref}[1]{\mbox{\rm{(\ref{#1})}}}    
\newcommand{\qn}{\frac{(n+1)(n+2)}{2}}   
\newcommand{\R}{\mbox{${\mathbb R}$}}           
\newcommand{\xb}{\mathbf{x}}  
\newcommand{\cb}{\mathbf{c}}  
\newcommand{\xh}{\hat{\xb}}  
\newcommand{\nx}{n_x}  
\newcommand{\iset}{\mathcal{X}} 
\newcommand{\tb}{\bm{\theta}} 
\newcommand{\nt}{n_{\theta}}  
\newcommand{\nd}{n_d}  
\newcommand{\eps}{\varepsilon} 
\newcommand{\epsb}{\bm{\eps}} 
\newcommand{\gb}{\mathbf{g}}  
\newcommand{\Hb}{\mathbf{H}}  
\newcommand{\cB} {\mbox{$\cal B$}} 
\newcommand{\Fb}{\mathbf{F}}  
\newcommand{\db}{\bm{\delta}} 
\newcommand{\cov}{\text{Cov}}
\newcommand{\var}{\text{Var}}
\newcommand{\algo}{\textsc{pound}er\textsc{s}}

\newcommand{\BdG}{\textsc{b}{\footnotesize d}\textsc{g}}
\newcommand{\PWscf}{\textsc{pw}scf}
\newcommand{\DFT}{\textsc{dft}}
\newcommand{\RMF}{\textsc{rmf}}
\newcommand{\SLDA}{\textsc{slda}}
\newcommand{\HO}{\textsc{ho}}
\newcommand{\THO}{\textsc{tho}}
\newcommand{\HF}{\textsc{hf}}
\newcommand{\BCS}{\textsc{bcs}}
\newcommand{\LN}{\textsc{ln}}
\newcommand{\EFA}{\textsc{efa}}
\newcommand{\HFODD}{\textsc{hfodd}}
\newcommand{\HFBTHO}{\textsc{hfbtho}}
\newcommand{\CC}{\textsc{cc}}
\newcommand{\CCSD}{\textsc{ccsd}}
\newcommand{\BLAS}{\textsc{blas}}
\newcommand{\LAPACK}{\textsc{lapack}}
\newcommand{\ATLAS}{\textsc{atlas}}
\newcommand{\GNU}{\textsc{gnu}}
\newcommand{\CECO}{\textsc{ceco}}
\newcommand{\UNEDFPRE}{\textsc{unedf}pre}
\newcommand{\UNEDFNB}{\textsc{unedf}nb}
\newcommand{\exclude}[1]{}
\newcommand{\mat}[1]{\mathbf{#1}}
\newcommand{\ket}[1]{|#1\rangle}
\newcommand{\bra}[1]{\langle#1|}
\newcommand{\norm}[1]{\lVert{#1}\rVert}
\newcommand{\braket}[1]{\mathinner{\langle{#1}\rangle}}{\catcode`\|=\active
  \gdef\Braket#1{\left<\mathcode`\|"8000\let|\bravert {#1}\right>}}
\newcommand{\bravert}{\egroup\,\vrule\,\bgroup}

\newcommand{\Ck}{C_\text{k}}
\newcommand{\enm}{E^\text{NM}}
\newcommand{\rhoc}{\rho_\text{c}}
\newcommand{\knm}{K^\text{NM}}

\newcommand{\asym}{a_\text{sym}^\text{NM}}
\newcommand{\lsym}{L_\text{sym}^\text{NM}}
\newcommand{\ksym}{\Delta K^\text{NM}}

\newcommand{\gras}[1]{\boldsymbol{#1}}
\newcommand{\capital}[1]{\mathscr{#1}}

\newcommand{\ba}{\begin{array}}
\newcommand{\ea}{\end{array}}
\newcommand{\disregard}[1]{}

\newcommand{\ali}{ali}

\newcommand{\tl}[1]{{\color{red}{#1}}}
\newcommand{\etl}{}
\newcommand{\tlfn}[1]{{\color{red}\footnote{\color{red}{TL: {#1}}}}}

\begin{abstract}
We carry out state-of-the-art optimization of a nuclear energy density of Skyrme type in the framework of the Hartree-Fock-Bogoliubov (HFB) theory. The particle-hole and particle-particle channels are optimized simultaneously, and the experimental data set includes both spherical and deformed nuclei. 
The new model-based, derivative-free optimization algorithm  used in this work has been 
found to be significantly better than standard 
 optimization methods in terms of reliability, speed, accuracy, and precision.
The resulting parameter set {\UNEDFPRE} results in good agreement with 
experimental masses, radii, and deformations and seems to be free of finite-size 
instabilities. An estimate of the reliability of the obtained parameterization is given, based on standard statistical methods.  We discuss new physics insights offered by the advanced covariance analysis.
\end{abstract}


\maketitle


\section{Introduction}
\label{Sec-introduction}

The goal of low-energy nuclear physics is to understand nuclei and how they react. This fascinating science problem is relevant to other fields and to a gamut of societal applications.  New vistas have been opened by experimental advances in the production of rare isotopes \cite{[IUPAP09]} and  
new theoretical approaches \cite{riatheory}  backed by unprecedented computing power  \cite{[Top50]}. The rapid experimental  developments have resulted in  a wealth of unique data from previously unexplored regions of the nuclear landscape. This situation poses a serious challenge  to  models of nuclear structure 
and  calls for their improved reliability and better-controlled extrapolability.

Theorists seek to formulate a coherent framework for  nuclear structure and reactions based on  a well-founded microscopic theory that would deliver maximum predictive power with well-quantified uncertainties. To this end, the steady increase in  computing power, currently crossing  the petaflop barrier, has been  beneficial. A paradigm for the new mode of nuclear theory is the SciDAC Universal Nuclear Energy Density Functional (UNEDF) project \cite{[Ber07]}, an example of the close alignment of the physics research with the necessary applied mathematics and computer science research. 

This study is the fruit of such a partnership, under UNEDF, in which physicists collaborate with mathematicians and computer scientists on a specific science challenge. Our long-term  goal in UNEDF is to develop a 
spectroscopic-quality theoretical framework rooted in the nuclear density functional theory (DFT)  \cite{[Ben03]}. In the first phase of the project, we have developed efficient DFT solvers for the self-consistent
Hartree-Fock-Bogoliubov (HFB) problem.  Various improvements that we have implemented to carry out
large-scale DFT calculations have been recently presented in  \cite{[Sto09],[Sch10]}. These improvements enable   
comprehensive mass-table calculations, including all even-even nuclei and many
different configurations in odd-even and odd-odd nuclei, in less than a day 
\cite{[Sto06],[Sto09a]}. 

The second phase of the project concerns the development and optimization of the nuclear energy  density functional (EDF). Since standard functionals are clearly too restrictive when one is aiming at a quantitative description \cite{[Ber05],[Kor08]}, the form of EDF needs to be improved. Novel functionals can be constructed from two- and three-nucleon interactions by using effective field theory and the density matrix expansion technique \cite{[Dru10],[Geb09],[Geb10],[Car10]}  and by  using constraints from ab initio calculations for very light nuclei and nuclear matter. They can also be obtained by enriching  density dependence  and adding higher gradient terms in a systematic way \cite{[Car10],[Car08],[Car09]}. 

Having determined the form of the EDF, one must still optimize the coupling constants of the underlying energy density (ED). Indeed, all energy functionals, irrespective of their theoretical foundations, rely on parameters that must be directly fitted to experimental data. It has been realized recently that high-performance computing can positively impact the optimization strategy. Historically, most nuclear ED parameterizations, such as  Skyrme or Gogny,  were obtained by a direct fit to selected experimental data from finite nuclei and various nuclear matter properties (NMPs). Observables commonly included in the fit are binding energies, proton radii, surface thickness, and/or single-particle (s.p.) energies of doubly closed-shell nuclei as well as NMPs (pseudo-observables) such as  energy per particle of infinite and semi-infinite nuclear matter, saturation density, or incompressibility. This is the case, for example, for the SLy4 parameterization of the Skyrme functional of \cite{[Cha95]}, which we take (somewhat arbitrarily) as a reference point in our study. The D1 and D1S parameterizations of the Gogny interaction have also been obtained in such a framework \cite{[Dec80],[Ber91]}. We refer to \cite{[Ben03],[Toi08],[Klu09]} for a more thorough discussion of  various fitting strategies and protocols.

In fact, very few examples of EDs are fitted to other types of data.
For Skyrme EDFs only, we mention the early attempt of the SkM*
parameterization \cite{[Bar82]}, which was adjusted semi-classically to account
for the fission barrier of $^{240}$Pu. The Brussels-Montreal  set of EDFs
has been optimized to data on deformed nuclei, although the actual fit is always performed with a
spherical code by using a multistep procedure. For example, in the early versions
MSk1-MSk6, the deformation energy of the ground-state configuration was used to
renormalize nuclear masses so that the optimization could proceed in spherical
symmetry \cite{[Ton00]}. Similarly, while in the later version HFB14, data on fission
barriers were used, the core part of the fitting procedure was carried out in spherical geometry
\cite{[Gor07]}. For SLy4 itself, several parameters were fixed at values
empirically expected to yield a correct description of giant resonance energy
centroids in random-phase approximation (RPA) calculations, although no such
calculation was included in the fit nor any quantitative check performed a
posteriori. 

The choice to restrict the  data set of observables  to those pertaining to nuclear matter and spherical nuclei has almost always been dictated by practical
considerations: the cost of performing huge  numbers of
deformed HFB calculations was 
deemed too high. It was
also rightly argued that the driving terms  of the EDF could be pinned down
by considering spherical nuclei only. With the need for more precision, however, 
the limitation to spherical nuclei and NMP is clearly not sufficient. The advent of supercomputers makes it possible to free ourselves from this restriction.

Specifically, the availability of supercomputers has two consequences. First, one can now include in the set of fit observables data corresponding to deformed nuclei, odd-mass systems, excited states, and so forth. More comprehensive data sets should better constrain the various channels of the energy functionals, for example, its deformation or spin-polarization properties. It might soon become possible to directly optimize symmetry-restored  EDFs \cite{[Dug10]}, either in a single-reference \cite{[She00a],[She02],[Sto07],[Dob09g]} or a multireference \cite{[Lac09],[Dug09],[Ben09]} framework.

In addition,  in our quest for improved EDFs, a key step is to understand various constraints imposed by experimental data on ED parameters and the resulting uncertainty margins. Early attempts to use statistical methods of linear-regression and error analysis \cite{[Fri86]} have been revived recently and applied to determine the correlations between ED parameters, parameter uncertainties, and the errors of calculated observables \cite{[Ber05],[Kor08],[Toi08],[Klu09],[Rei10]}. This approach is essential for providing predictive capability and extrapolability and for estimating  the theoretical uncertainties. 

The purpose of this work is to revisit the problem of Skyrme ED optimization by (i) removing some of the previous limitations with the help of modern computational resources, and (ii) applying regression diagnostics methods on the resulting parameterization. To these ends, we perform functional optimization with a model-based method that is particularly adapted to costly function evaluations, such as when the objective function contains the result of hundreds of symmetry-unrestricted HFB calculations. In our model study, we focus on nuclear masses and radii, with a bias
toward heavy nuclei. The final ED parameterization is subjected to a fully fledged correlation and sensitivity analysis. While we do not claim to have found an end-all parameterization of the  Skyrme EDF, we believe that the set of techniques  we have applied in this study can  pave the way to a universal nuclear EDF of spectroscopic quality.

The paper is organized as follows. In Sec.~\ref{Sec-framework} we briefly present the DFT framework used, in particular various parameterizations of the Skyrme EDF and their relations to nuclear matter properties. We also discuss the choice of experimental observables. Section \ref{Sec-optimization} presents the specific model-based algorithm used in this work and contains all the technical information related to large-scale  HFB calculations. Results are discussed in Sec. \ref{Sec-results}. Section~\ref{conclusions} contains the conclusions of this work.


\section{Theoretical Framework}
\label{Sec-framework}

This section recalls the features of the Skyrme-DFT theory that are relevant to the optimization problem. A  detailed presentation of the theory itself can be found in, for example, \cite{(Rin80),[Ben03],[Per04]} and references therein. The main focus of the following discussion is on  various parameterizations of the Skyrme EDF and the selection of experimental observables chosen to constrain ED parameters.


\subsection{Time-Even Skyrme Energy Density Functional}
\label{Subsec-skyrme}

In nuclear DFT, the total energy of the nucleus is given by
\begin{equation}
E =  \int 
    \mathcal{H}(\gras{r}) d^{3}\gras{r} 
\end{equation}
where $\mathcal{H}$ is the local energy density that is supposed to be
a real, scalar, time-even, and isoscalar function of local densities 
and their derivatives. The Skyrme ED can be decomposed into the kinetic term, 
interaction ED $\chi$, pairing ED, Coulomb term, and additional corrections, 
such as the center-of-mass term. For the kinetic energy term, we set 
$\hbar^{2}/2m$=20.73553\,MeV\,fm$^{2}$. The Coulomb Hartree term is 
calculated exactly, while the exchange term is computed by the Slater 
approximation. The contribution from the center-of-mass correction 
has the same structure as the kinetic term and leads to a renormalization 
of the nucleon mass $1/m\rightarrow (1/m)[1-1/A]$. All these prescriptions 
follow the SLy4 parameterization. 

The interaction ED can be further decomposed into $\chi=\chi_0 + \chi_1$, 
with
\begin{eqnarray}
\chi_t(\gras{r})&=& C_t^{\rho\rho} \rho_t^2 
  + C_t^{\rho\tau} \rho_t\tau_t +  C_t^{J^2} \bm{J}_t^2
\nonumber \\ &&
  + C_t^{\rho\Delta\rho} \rho_t\Delta\rho_t \
  + C_t^{\rho \nabla J}  \rho_t\bm{\nabla}\cdot\bm{J}_t,
\label{UED}
\end{eqnarray}
where the isospin index $t$ labels isoscalar ($t$=0) and isovector ($t$=1)
densities. Since in this work we limit the discussion to even-even nuclei, 
the terms involving spin, spin-kinetic, and current densities 
\cite{[Per04],[Eng75],[Ben03]} are absent. The coupling constants 
$C_t^{\rho\rho}$ contain an additional dependence on the isoscalar 
density of the form
\begin{equation}
C_t^{\rho\rho} =  C_{t0}^{\rho\rho}
  + C_{t{\rm D}}^{\rho\rho}~ \rho_0^\gamma. 
\label{crramp}
\end{equation}

The standard Skyrme interaction ED therefore contains 13 independent 
parameters:
\begin{equation}
\{C_{t0}^{\rho\rho}, C_{t{\rm D}}^{\rho\rho}, C_t^{\rho\Delta\rho}, 
  C_t^{\rho\tau},    C_t^{J^2},               C_t^{\rho\nabla J} \}_{t=0,1}\ 
  \text{and}\ \gamma.
\label{c-parameters}
\end{equation}
When dealing with the Skyrme interaction EDF (i.e., the functional that 
originates from the Skyrme interaction), the coupling constants 
(\ref{c-parameters}) are uniquely related to the well-known 
$(t,x)$-parameterization of the Skyrme interaction
\begin{equation}
\{t_0, t_1, t_2, t_3, 
  x_0, x_1, x_2, x_3, 
  t_\text{o}, t_\text{e}, b_4, b_4', \gamma \}.
\label{t-parameters}
\end{equation}
The equations connecting the $C$- and $(t,x)$-parameterization can be 
found, for example, in \cite{[Per04]}.

In this study, nucleonic  superconductivity  is described by the pairing ED: 
\begin{equation}
\breve{\chi}(\gras{r}) = \sum_{q=n,p}
{V^{q}_0\over 2}
\left[ 1 - \frac{1}{2}\frac{\rho(\gras{r})}{\rho_0} \right]\breve\rho^2(\gras{r}) ,
\label{vpair}
\end{equation}
where  $\breve\rho$ is the local pairing density and  $\rho_{0}$=0.16\,fm$^{-3}$ 
(mixed-pairing prescription \cite{[Dob02c]}).


\subsection{Nuclear Matter Properties and Skyrme Energy Density Parameterizations}
\label{Subsec-NM}

The $(t,x)$ and $C$-representations are natural parameterizations of the Skyrme EDF, 
the former in terms of an effective, density-dependent two-body interaction and the 
latter as a general functional of the density. However, these representations do not 
provide a straightforward connection to physical observables; hence, it is not 
immediately obvious  what the search range for these parameters should be. It is 
therefore advantageous to relate them to fundamental properties of symmetric and 
asymmetric homogeneous nuclear matter, which have a clear physical interpretation 
and the range of which is known \cite{[Cha97],[Sto07b],[Klu09]}.

The starting point in the discussion of NMPs is  the equation of state (EOS) of 
the infinite homogeneous nuclear matter: $E/A=W(\rho_n,\rho_p)$. The Coulomb energy 
is disregarded, all gradient terms vanish, and the kinetic energy density is 
replaced by its Thomas-Fermi expression. Assuming an unpolarized system, one 
can also ignore terms involving time-odd spin densities and currents. 

The expansion of $W(\rho_n,\rho_p)$ around the equilibrium density $\rhoc$ and 
$I = 0$ can be written as
\begin{equation}
W(\rho_n, \rho_p) = W(\rho_0,I) = W(\rho_0) + S_2(\rho_0) I^2 +
\mathcal{O}(I^4),
\end{equation}
where 
$I = \rho_1/\rho_0 =
(\rho_{n}-\rho_{p})/\rho_0$ is the relative neutron excess,
$\rho_0 = \rho_{n} + \rho_{p}$, $\rho_1 = \rho_{n} - \rho_{p}$, 
\begin{equation}
 W(\rho_0)    = 
\frac{\enm}{A}+\frac{P^\text{NM}}{ \rhoc^2 } \left( \rho_0-\rhoc \right) 
  + \frac{\knm}{18 \rhoc^2} \left( \rho_0-\rhoc \right)^2,
\end{equation}  
and
\begin{eqnarray}
 S_2(\rho_0)  & = &
\asym + \frac{\lsym}{3 \rhoc} \left( \rho_0 - \rhoc \right) 
\nonumber \\
 & & + \frac{\ksym}{18 \rhoc^2} \left( \rho_0 - \rhoc \right)^2.
\label{NMT}
\end{eqnarray}
In these equations, $\enm/A$ stands for the total energy per nucleon at 
equilibrium, $P^\text{NM}$ represents the nucleonic pressure, $\knm$ is the  
nuclear matter incompressibility, $\asym$ is the symmetry energy coefficient, 
$\lsym$  represents the density dependence of the symmetry energy, and $\ksym$ is a correction 
to the incompressibility.


\subsubsection{Symmetric nuclear matter}
\label{Subsubsec-SNM}

In the regime of symmetric nuclear matter (SNM), $\rho_\text{n} =
\rho_\text{p} = \rho_0/2$ and $I$=0, which eliminates all isovector terms. The
isoscalar kinetic energy density is
\begin{equation}
 \tau_0 = \Ck \rho_0^{5/3}, 
  ~~~ \Ck= \frac{3}{5} \left(\frac{3 \pi^2}{2}\right)^{2/3}.
\label{tauTF}
\end{equation}
The nuclear matter saturation curve $W(\rho_0)$ is expected to have the
following  properties:
\begin{eqnarray}
\rhoc  ~\approx &  0.16 ~ \text{fm}^{-3},& ~ \\
 P^\text{NM} ~=& 
  \left. \rho^2 \displaystyle\frac{d W(\rho_0)}{d\rho_0} \right|_{\rho_0=\rhoc}
    &=~  0,
\\
  \frac{\enm}{A} ~=& 
  W(\rhoc) & \approx ~  -16~ \text{MeV}.
\label{symNMexpbegin}
\end{eqnarray}
The value of the incompressibility modulus is related to the centroid energies 
of giant isoscalar monopole resonances in isospin-symmetric nuclei \cite{[Pie09]} 
and is expected to be \cite{[Bla80],[Col04]}
\begin{equation}
 \knm = 
  \left. 9 \rho_0^2 \frac{d^2 W(\rho_0)}{d\rho_0^2}
    \right|_{\rho_0=\rhoc} \approx 220\pm 10~\text{MeV},
\end{equation}
with a strong preference for 230\,MeV \cite{[Tod05]}.
Another important NMP, entering the SNM EOS indirectly, is the
isoscalar effective mass
\begin{equation}
 M_s^{* -1} =
  \frac{2m}{\hbar^2} \left. \frac{d E}{d\tau_0} \right|_{\rho_0=\rhoc},
\label{symNMexpend}
\end{equation}
which quantifies the momentum-dependence of the mean field and drives
the density of the s.p. spectrum. An appropriate value for a
fit to experimental s.p. energies is $M_s^{*}=1$ \cite{[Bro98]}, while ab initio
calculations performed at the Brueckner-Hartree-Fock level in INM suggest a
slightly lower value for the Landau (Fermi-level) effective mass extracted from
the on-shell s.p. spectrum \cite{[Zuo99],[Zuo02],[van05],[Heb09]}.
Mass fits also seem to favor a value close to unity, although 
significant freedom exists \cite{[Gor03]}.

The SNM EOS expressed in terms of the coupling constants of the
Skyrme EDF is
\begin{eqnarray}
W(\rho_0) & = &
 \left(\frac{\hbar^2}{2m} +
       C_0^{\rho\tau} \rho_0 \right) \Ck \rho_0^{2/3} 
\nonumber \\ & & 
+\left( C_{00}^{\rho\rho} 
+ C_{0D}^{\rho\rho} \rho_0^\gamma \right)  \rho_0.
\label{wC}
\end{eqnarray}

Computing the quantities (\ref{symNMexpbegin}-\ref{symNMexpend}) 
using (\ref{wC}) allows us to express the coupling constants 
$C_{00}^{\rho\rho}$, $C_{0D}^{\rho\rho}$, $C_0^{\rho\tau}$ and 
the power $\gamma$ in terms of $\enm/A$, $P^\text{NM} = 0$, $\knm$
and $M_s^{* -1}$. The resulting expressions are \cite{[Cha97]} as follows:
\begin{eqnarray}
C_{00}^{\rho\rho} & = & 
\frac{1}{3 \gamma \rhoc} \left\{\tfrac{\hbar^2}{2m} 
\left[ \left( 2 - 3 \gamma \right) M_s^{* -1} - 3 \right] \tau_c
\right. \nonumber \\ & & \left.
+ 3\left(1 + \gamma\right) \tfrac{\enm}{A}\right\}, \\
C_{0D}^{\rho\rho} & = & 
\frac{1}{3 \gamma \rhoc^{1 +\gamma} } \left[
\tfrac{\hbar^2}{2m} \left(3 - 2M_s^{* -1}\right) \tau_c
-3\tfrac{\enm}{A} \right],\\
C_0^{\rho\tau} & = &
 \frac{\hbar^2}{2m} \left(M_s^{* -1}-1\right)\frac{1}{\rhoc},
 \\
\gamma & = &
 \frac{\tfrac{\hbar^2}{2m} \left(4 M_s^{* -1} - 3\right) \tau_c 
  - \knm - 9 \frac{\enm}{A}}{\frac{\hbar^2}{2m}
\left(6 M_s^{* -1}-9\right) \tau_c + 9 \tfrac{\enm}{A}},
\label{CfromSNM}
\end{eqnarray}
where $\tau_{c} = \Ck \rho_{c}^{5/3}$.


\subsubsection{Asymmetric nuclear matter}
\label{Subsubsec-ANM}

In asymmetric nuclear matter (ANM), neutron and proton densities are different,  
and isovector terms are nonzero. The local and kinetic energy densities are
\begin{eqnarray}
\rho_1 &=& I  \rho_0, \\
\tau_0 &=&   \Ck \rho_0^{2/3} F_+(I), \\
\tau_1 &=&   \Ck \rho_0^{2/3} F_-(I),
\label{tauTFa} \\
F_\pm(I) &=& \frac{1}{2} \left[ (1 + I)^{5/3} \pm (1 - I)^{5/3} \right].
\end{eqnarray}
The nuclear matter EOS $W(I,\rho)$ now depends on the relative neutron excess $I$. 
The most important parameter characterizing the isospin dependence of the ANM EOS is
the symmetry energy at saturation density,
\begin{equation}
S_2(\rhoc) = \asym = \left. \frac{1}{2} 
  \frac{d^2 W(\rho_0,I)}{d I^2}
  \right|_{\begin{array}{l} {\rho_0=\rhoc}\\{I=0}\end{array}}.
\label{asymNMexpbegin}
\end{equation}
The value of $S_2(\rhoc)$ varies from 28 to 36~MeV among EDFs extrapolated to nuclear
matter \cite{[Sto07b],[Li08]}. It is understood \cite{[Ton84],[Rei06]} that nuclear 
masses constrain a combination of the symmetry- and surface-symmetry energy parameters 
in a given EDF, and this fact explains the large spread of values. 

The variation of the density-dependent symmetry energy $S_2$ with $\rho_0$ is
usually parameterized through
\begin{equation}
 \lsym = 3 \rhoc 
  \left. \frac{d S_2(\rho_0)}{d\rho_0} \right|_{\rho_0=\rhoc},
\end{equation}
the value of which appears correlated with the thickness of neutron skins in 
asymmetric nuclei (see \cite{[Rei10]} and references therein). An empirical 
determination of this parameter yields $\lsym=80\pm30$\,MeV \cite{[Che05],[Che09]}. One now introduces
\begin{equation}
 \ksym = 9 \rhoc^2 
  \left. \frac{d^2 S_2(\rho_0)}{d^2\rho_0} \right|_{\rho_0=\rhoc},
\end{equation}
which affects the incompressibility of the ANM and thus the isoscalar monopole 
resonance energies in neutron-rich nuclei \cite{[Pie09]}. For the SLy4 EDF, 
the values of the last two parameters were determined by the fit to the 
neutron matter EOS. We let these quantities be constrained by our 
experimental data set. We will see whether these data leave enough freedom to 
apply additional constraints in the regression analysis. 

The momentum dependence of the mean field is also affected by
isospin: neutron and proton effective masses are different in asymmetric
matter \cite{[Les07]}, an effect quantified by the isovector effective mass 
\begin{equation}
 M_v^{* -1} = M_s^{* -1}
  - \frac{2m}{\hbar^2} \left. \frac{d E}{d\tau_1}
   \right|_{\begin{array}{l}{\rho_0=\rhoc}\\{I=0}\end{array}}.
\label{asymNMexpend}
\end{equation}

The EOS of homogeneous asymmetric nuclear matter can be written as
\begin{multline}
W(I,\rho_{0}) =
\left(\frac{\hbar^2}{2m}  + C_0^{\rho \tau} \rho\right)
\Ck\, \rho_0^{2/3}\, F_+(I) \\
+ C_1^{\rho \tau} \Ck\, \rho_0^{5/3}\, I F_-(I)  \\
+ \left[ C_{00}^{\rho\rho} +  C_{0D}^{\rho\rho} \rho_0^\gamma 
+ I^2 \left(C_{10}^{\rho\rho} + C_{1D}^{\rho\rho} \rho_0^\gamma\right)\right]
\rho_0.
\label{wCa}
\end{multline}
Just as for SNM, we compute the quantities
(\ref{asymNMexpbegin}-\ref{asymNMexpend}) from (\ref{wCa}) and obtain an expression for $C_{10}^{\rho\rho}$, $C_{1D}^{\rho\rho}$, 
and $C_1^{\rho\tau}$ \cite{[Cha97]}:
\begin{eqnarray}
C_1^{\rho\tau}    & = & 
C_0^\tau - \frac{\hbar^2}{2m} \left(M_{v}^{* -1}-1\right)\frac{1}{\rhoc},
\\
C_{10}^{\rho\rho} & = & \left.
\frac{1}{27 \gamma \rhoc} \right[ 
27 \left( 1 + \gamma \right) \asym - 9 \lsym 
\nonumber \\ & & 
+5\tau_c  \left(2-3 \gamma\right)
\left(C_0^\tau + 3 C_1^\tau\right)\rhoc
\nonumber \\ & & \left. 
-5\tau_c  \left(1 + 3 \gamma\right) \frac{\hbar^2}{2m} \right] ,
\\
C_{1D}^{\rho\rho} & = & 
  \left. \frac{1}{27 \gamma  \rhoc^{\gamma+1}}
  \right[-27 \asym + 9 \lsym
\nonumber \\
 & & +\left. 5 \left(\frac{\hbar^2}{2m} 
  - 2 \rhoc \left(C_0^\tau + 3 C_1^\tau\right)\right) \tau_c \right].
\label{CfromANM}
\end{eqnarray}

Using relations (\ref{CfromSNM})-(\ref{CfromANM}), we express 7 of the 
original 13  parameters (\ref{c-parameters}) of the Skyrme EDF as functions of 
nuclear matter properties. The remaining 6 are not known exactly and should 
therefore not be used as rigid constraints \cite{[Klu09]}. However, 
the expected values of all these NMPs are 
sufficient to provide well-defined intervals of variation during the optimization process. The 6 remaining 
coupling constants are the isoscalar and isovector $C_{t}^{\rho\Delta\rho}$, 
spin-orbit $C_{t}^{\rho\nabla J}$, and tensor $C_{t}^{J^{2}}$ terms. 
Consequently, the 
Skyrme EDF   depends on the following 13 parameters:
\begin{multline}
\left\{
\rhoc, \enm/A, M_s^{*}, \knm, \asym, \lsym, 
\right. \\
\left.
  M_v^{*}, C_{0}^{\rho\Delta\rho}, C_{1}^{\rho\Delta\rho}, 
  C_{0}^{\rho\nabla J}, C_{1}^{\rho\nabla J},
  C_{0}^{J^{2}}, C_{1}^{J^{2}}
\right\}.
\end{multline}


\subsection{Fit Observables}
\label{Subsec-experimental}

To calibrate the EDF, we selected a pool of fit observables that constitute 
the UNEDF experimental database \cite{[Dat09]}. The purpose of the database 
is to provide a standard and comprehensive set of experimental data that can  
be used to systematically optimize  EDFs. Since we wish to provide, together 
with the optimized set of parameters, a measure of its intrinsic quality via 
the error and sensitivity analysis, for every observable an error bar should 
also be defined. We organized our database into three major categories --
spherical, deformed, and symmetry-unrestricted -- which reflect the 
level of symmetry-breaking of the underlying EDF and thereby the complexity 
of its numerical implementation. More details can be found in \cite{[Dat09]}.

\begin{figure}
\includegraphics[width=0.98\columnwidth]{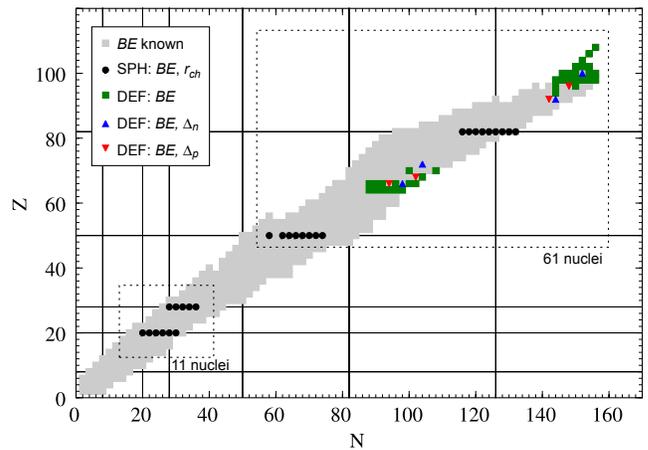}
\caption{(color online) Experimental  set of fit observables used in this work. 
The set contains data for 11 nuclei with $A\!\!<$66 and 61 nuclei with $A\!\!>$106.}
\label{fig-data}
\end{figure}

The focus of this  work is on a well-controlled optimization methodology, and 
the emphasis is on global nuclear properties such as masses and proton radii. 
Our functional is therefore restricted to time-even densities, and only 
spherical or axially deformed nuclei are considered. The chosen observables 
embrace data for 72 nuclei, which are proven to allow a reasonable DFT description. 
The selected experimental data set is presented in Fig.~\ref{fig-data}. 
As can be seen, the emphasis is on the heavy nuclei. Indeed, there are only 11 
nuclei with $A\!\!<$66 in our data set. Below, we give a detailed description 
of the  set of fit observables used in this work.

\begin{table}[h]
\begin{center}
\caption{Nuclear binding energies (in MeV; the electronic energy correction 
has been subtracted) \cite{[Aud03]} for the 44 deformed nuclei selected in 
this work. The column marked ``\#'' is the data point number.
        }
\begin{ruledtabular}
\begin{tabular}{cccccccc}
\# & Z & N & E  & \# & Z & N & E \\
\hline
 1 & 108 & 156 & -1925.697   &   23 &  94 & 144 & -1800.523 \\
 2 & 106 & 154 & -1908.038   &   24 &  92 & 144 & -1789.701 \\
 3 & 104 & 152 & -1889.709   &   25 &  92 & 142 & -1777.858 \\
 4 & 102 & 154 & -1897.729   &   26 &  90 & 142 & -1766.015 \\
 5 & 102 & 152 & -1884.685   &   27 &  72 & 104 & -1418.407 \\
 6 & 102 & 150 & -1870.386   &   28 &  70 & 108 & -1431.260 \\
 7 & 100 & 156 & -1901.673   &   29 &  70 & 100 & -1377.760 \\
 8 & 100 & 154 & -1890.112   &   30 &  68 & 104 & -1391.213 \\
 9 & 100 & 152 & -1878.056   &   31 &  68 & 102 & -1378.695 \\
10 & 100 & 150 & -1864.657   &   32 &  66 & 102 & -1362.591 \\
11 & 100 & 148 & -1850.682   &   33 &  66 & 100 & -1350.474 \\
12 & 100 & 146 & -1836.305   &   34 &  66 &  98 & -1337.714 \\
13 &  98 & 156 & -1891.281   &   35 &  66 &  96 & -1323.785 \\
14 &  98 & 154 & -1880.445   &   36 &  66 &  94 & -1309.134 \\
15 &  98 & 152 & -1869.165   &   37 &  66 &  92 & -1293.725 \\
16 &  98 & 150 & -1856.954   &   38 &  66 &  90 & -1277.701 \\
17 &  98 & 148 & -1843.959   &   39 &  64 &  98 & -1321.473 \\
18 &  98 & 146 & -1830.429   &   40 &  64 &  96 & -1308.992 \\
19 &  98 & 144 & -1816.428   &   41 &  64 &  94 & -1295.597 \\
20 &  96 & 150 & -1847.037   &   42 &  64 &  92 & -1281.300 \\
21 &  96 & 148 & -1835.059   &   43 &  64 &  90 & -1266.329 \\
22 &  96 & 144 & -1809.502   &   44 &  64 &  88 & -1251.187 \\
\end{tabular}
\end{ruledtabular}
\label{table:edef}
\end{center}
\end{table}

\subsubsection{Deformed nuclei} 

In our optimization, we considered binding energies of 44 well-deformed 
even-even nuclei shown in Table~\ref{table:edef}. Candidates were selected 
from an HFB mass-table calculation with the SLy4 parameterization requiring 
that their ground-state equilibrium deformation be greater than $|\beta|$=0.25. 
Since the majority of atomic nuclei are deformed in their ground states, by 
including binding energies of deformed systems in the database, one hopes to 
better probe the surface properties of the EDF.


\subsubsection{Spherical nuclei} 

Table~\ref{table:esphe} lists the nuclear masses of a selected set of 28 
spherical nuclei considered in the fit. In these nuclei, correlations 
beyond mean-field are expected to be relatively constant \cite{[Klu09]}. 
Since the list includes doubly magic nuclei, it should provide strong 
constraints, as these nuclei tend to deviate from global mass trends 
\cite{[Ber05]}. Moreover, the masses of $^{40}$Ca, $^{48}$Ca, and $^{56}$Ni 
help constrain the spin-orbit term \cite{[Bei75],[Les07],[Zal08]}.
All the masses of spherical and deformed nuclei given in 
Tables~\ref{table:edef}-\ref{table:esphe} have been corrected for the electronic 
binding energy. The nuclear binding energy $E_{\text{nuc}}(Z,N)$ is given by 
\begin{equation}
E_{\text{nuc}}(Z,N) = E_{\text{ato}}(Z,N) - E_{\text{el}},
\end{equation}
where $E_{\text{ato}}(Z,N)$ is the atomic binding energy and 
$E_{\text{el}} = -1.433 \times 10^{-5} Z^{2.39}\,{\rm MeV}$.

\begin{table}[h]
\begin{center}
\caption{Nuclear binding energies (in MeV; the electronic energy correction 
has been subtracted) \cite{[Aud95]} for 28 spherical nuclei selected in this 
work. The column marked ``\#'' is the data point number.
        }
\begin{ruledtabular}
\begin{tabular}{cccccccc}
\# & Z  & N   & E &  \# & Z  & N   & E    \\
\hline
45 & 82 & 132 & -1662.762    &   59 & 50 &  64 & -971.406  \\
46 & 82 & 130 & -1653.988    &   60 & 50 &  62 & -953.335  \\
47 & 82 & 128 & -1645.030    &   61 & 50 &  58 & -914.424  \\
48 & 82 & 126 & -1635.909    &   62 & 28 &  36 & -561.714  \\
49 & 82 & 124 & -1621.803    &   63 & 28 &  34 & -545.217  \\
50 & 82 & 122 & -1606.984    &   64 & 28 &  32 & -526.801  \\
51 & 82 & 120 & -1591.666    &   65 & 28 &  30 & -506.459  \\
52 & 82 & 118 & -1575.833    &   66 & 28 &  28 & -483.949  \\
53 & 82 & 116 & -1559.483    &   67 & 20 &  30 & -427.473  \\
54 & 50 &  74 & -1049.835    &   68 & 20 &  28 & -415.972  \\
55 & 50 &  72 & -1035.365    &   69 & 20 &  26 & -398.751  \\
56 & 50 &  70 & -1020.375    &   70 & 20 &  24 & -380.942  \\
57 & 50 &  68 & -1004.785    &   71 & 20 &  22 & -361.877  \\
58 & 50 &  66 &  -988.535    &   72 & 20 &  20 & -342.033  \\
\end{tabular}
\end{ruledtabular}
\label{table:esphe}
\end{center}
\end{table}

For the same 28 spherical nuclei, we also consider the proton rms point radius 
$\langle R_{\text{p}}^{2} \rangle$, which we extract from the charge radius 
$\langle R_{\text{ch}}^{2} \rangle$ of \cite{[Fri95]} using  the 
standard relation:
\begin{equation}
\langle R_{\text{ch}}^{2} \rangle = 
\langle R_{\text{p}}^{2} \rangle + 
\langle r_{\text{p}}^{2} \rangle + 
\frac{N}{Z}\langle r_{\text{n}}^{2} \rangle,
\end{equation}
where the proton charge radius, $\sqrt{\langle r_{\text{p}}^{2} \rangle} = 0.877$~fm,  and 
the neutron charge radius, $\langle r_{\text{n}}^{2} \rangle = -0.1161$~fm$^{2}$, were  taken from \cite{[Ams08]}. The values of proton radii used in this work are listed in Table~\ref{table:radius}.

\begin{table}[h]
\begin{center}
\caption{Proton rms radii (in fm) \cite{[Fri95]} for the 28 spherical nuclei 
selected in this work. The column marked ``\#'' is the data point number.
        }
\begin{ruledtabular}
\begin{tabular}{cccccccc}
\# & Z  & N   & $r_p$  & \# & Z  & N   & $r_p$   \\
\hline
73 & 82 & 132 & 5.506   &   87 & 50 & 64 & 4.542 \\
74 & 82 & 130 & 5.488   &   88 & 50 & 62 & 4.527 \\
75 & 82 & 128 & 5.469   &   89 & 50 & 58 & 4.492 \\
76 & 82 & 126 & 5.450   &   90 & 28 & 36 & 3.787 \\
77 & 82 & 124 & 5.439   &   91 & 28 & 34 & 3.765 \\
78 & 82 & 122 & 5.428   &   92 & 28 & 32 & 3.733 \\
79 & 82 & 120 & 5.418   &   93 & 28 & 30 & 3.689 \\
80 & 82 & 118 & 5.403   &   94 & 28 & 28 & 3.661 \\
81 & 82 & 116 & 5.394   &   95 & 20 & 30 & 3.437 \\
82 & 50 &  74 & 4.609   &   96 & 20 & 28 & 3.390 \\
83 & 50 &  72 & 4.598   &   97 & 20 & 26 & 3.412 \\
84 & 50 &  70 & 4.586   &   98 & 20 & 24 & 3.432 \\
85 & 50 &  68 & 4.573   &   99 & 20 & 22 & 3.420 \\
86 & 50 &  66 & 4.558   &  100 & 20 & 20 & 3.382 \\
\end{tabular}
\end{ruledtabular}
\label{table:radius}
\end{center}
\end{table}


\subsubsection{Pairing} 

Since the particle-hole and particle-particle channels cannot easily be 
disentangled, we must also include observables that will help us pin down 
the magnitude of pairing correlations. Usually, the pairing part of the EDF 
is constrained by considering the odd-even staggering (OES) of binding energy 
(see \cite{[Ber09a]} for a recent survey). Additional constraints on the 
pairing ED may be imposed by taking calculated pairing gaps in symmetric 
nuclear matter and neutron matter \cite{[Cao06]}. This strategy has been 
adopted by the Brussels-Montreal group in their most recent model HFB-17 
\cite{[Gor09]}.

\begin{table}[h]
\begin{center}
\caption{Values of the neutron and proton average odd-even mass staggering (in MeV)
considered in this work. The column marked ``\#'' is the data point number.
        }
\begin{ruledtabular}
\begin{tabular}{cccc|cccc}
 \multicolumn{4}{c}{Neutrons} & \multicolumn{4}{c}{Protons}\\
\# & Z & N & $\tilde{\Delta}_{n}^{(3)}$  & \# & Z & N &  $\tilde{\Delta}_{p}^{(3)}$ \\
\hline
101 & 100 & 152 & 0.515   &   105 & 96 & 148 & 0.566 \\
102 &  92 & 144 & 0.569   &   106 & 92 & 142 & 0.606 \\
103 &  72 & 104 & 0.675   &   107 & 68 & 102 & 0.504 \\
104 &  66 &  98 & 0.679   &   108 & 66 &  94 & 0.728 \\
\end{tabular}
\end{ruledtabular}
\label{table:OES}
\end{center}
\end{table}

In this work, we constrain pairing EDF by means of the OES defined by a 3-point formula $\Delta^{(3)}$ \cite{[Sat98a],[Ber09a]}. As customary, 
the theoretical result for even particle number $N$ is compared with the 
experimental $\Delta^{(3)}$ for $N + 1$ \cite{[Dob01fw]}. We took four values 
of $\Delta^{(3)}$ for neutrons and four for protons; see Table~\ref{table:OES}. 
All these nuclei belong to the deformed set of Table~\ref{table:edef}. Our choice 
has been motivated by the observation that fitting pairing properties in spherical 
systems, where the level density is much greater, may lead to an underestimation 
of the overall pairing strength \cite{[Ber09a]}. 

With the fairly simple pairing ED (\ref{vpair}) that we use, it is not essential 
to require very high precision for the OES. For that reason, in order to be free 
from local fluctuations, we chose in each even-even nucleus the average over 
the two even-odd or odd-even isotopes: $\tilde{\Delta}_{n}^{(3)}(N) = [ \Delta^{(3)}(N-1)
+ \Delta^{(3)}(N+1) ]/2$. Including average values of $\tilde{\Delta}^{(3)}$ in our 
data set ensures that the magnitude of pairing correlations is correct and remains 
such throughout the fitting procedure. Theoretical OES values have been computed from 
the average HFB pairing gap \cite{[Dob84],[Dob95c]}.


\section{Optimization Algorithm}
\label{Sec-optimization}

This section briefly presents the new algorithm used in our optimization. 
We refer to it by the acronym {\algo}, standing for Practical Optimization 
Using No Derivatives (for Squares). We also provide the numerical parameters 
used in the HFB calculations, and we give the characteristics of the objective 
function used in the optimization.


\subsection{Derivative-Free Optimization Method}
\label{Subsec-model}

To outline our algorithm, we adopt the following notation. We denote the set 
of parameters/coupling constants of the Skyrme EDF to be fitted by $\xb\in \R^{\nx}$, 
where $n_x$ is the number of coupling constants of components $x_{k}$ to fit.
We define a composite fit function made of $D_T$ different types of data: 
nuclear masses, proton radii, and so on. The number $n_i$ of data points for a given 
type $i$ may vary; for example, we have more masses than rms radii. The output 
of the calculation for type $i$ is denoted by $s_{i,j}(\xb)$ for nucleus $j$
and obviously depends on the parameterization of the functional, that is, the
vector $\xb\in \R^{\nx}$. For type $i$ and nucleus $j$, the experimental value
of a given observable is denoted $d_{i,j}$.

While many objectives are possible, we minimize the weighted sum of squared errors
\begin{equation}
\chi^2(\xb)=\frac{1}{n_d-\nx} \sum \limits_{i=1}^{D_T} \sum \limits_{j=1}^{n_{i}}
\left(\frac{ s_{i,j}(\xb)-d_{i,j} } { w_i } \right)^2,
\label{eq:chi2}
\end{equation}
where $n_d=\sum_{i=1}^{D_T} n_{i}$ denotes the total number of data points being
fit. The weights $w_i>0$ render the type $i$ difference dimensionless and
are chosen to balance the goals of fitting different observable types
simultaneously.

The objective \Ref{eq:chi2} is a special case of the nonlinear least squares  function
\begin{equation}
f(\xb) = \frac{1}{2} \sum_{i=1}^{n_d} F_i(\xb)^2 =\frac{1}{2} \|\Fb(\xb)\|^2,
 \label{eq:nls}
\end{equation}
where the function $\Fb:\R^{\nx}\rightarrow \R^{n_{d}}$ yields the vector of
reduced errors. Most optimization approaches to minimizing \Ref{eq:nls} are
based on Newton's method, whereby $f$ is replaced by its second-order expansion
\begin{multline}
f(\xb+\db) \approx f(\xb) + \db^TJ(\xb)^T \mathbf{F}(\xb) \\
+\frac{1}{2}\db^T\left( J(\xb)^T J(\xb) + \sum_{i=1}^{n_d} F_i(\xb) \nabla^2 F_i(\xb)\right) \db,
 \label{eq:newton}
\end{multline}
where $\boldsymbol{\delta}\in\mathbb{R}^{n_{x}}$ and $J(\xb)$ is the Jacobian matrix
$J(\xb) = [\nabla F_1(\xb) , \cdots , \nabla F_{n_d}(\xb)]^T$.

In the problem at hand (and many others), the derivatives of $F_i(\xb)$ with
respect to $\xb$, $\boldsymbol{\nabla} F_i(\xb)$ and $\nabla^2 F_i(\xb)$, 
exist for virtually all $\xb$, but their calculation for use in the optimization 
is impractical. Indeed, although derivatives of binding energies can be obtained 
through the Feynman-Hellman theorem, other observables such as radii would require 
the use of perturbation theory or a cumbersome and potentially imprecise calculation 
of numerical differences. In such a case, the optimization algorithm  must
be \emph{derivative-free}, relying only on the function value outputs
$\mathbf{F}(\xb)$. Popular algorithms in this setting include the Nelder-Mead
(N-M) method and other direct search algorithms \cite{[Kol03]} and genetic algorithms
and other heuristics \cite{(Gol89)}. However, a recent benchmarking study
\cite{[Mor09]} found that methods that form a smooth approximation model of the
objective in order to exploit the smoothness and structure of the objective may
be able to obtain better solutions in fewer evaluations.

In the case of nonlinear least squares, we follow the approach of forming a 
quadratic model for each component,
\begin{equation}
 q_i(\xb+\db) = F_i(\xb) + \db^T \gb_i + \frac{1}{2} \db^T \Hb_i \db,
\end{equation}
with $\gb_i$ and $\Hb_i=\Hb_i^T$ playing the role of the unknown derivatives
$\boldsymbol{\nabla} F_i(\xb)$ and $\nabla^2 F_i(\xb)$, respectively. We obtain 
the model parameters $\gb_i$ and $\Hb_i$ by requiring that the model $q_i$ agree 
with the true function $F_i$ on a set $\iset$ of $\xb$ values at which $F_i$ is known.
Mathematically, these parameters are solutions to the convex quadratic program
\begin{equation}
 \min_{\gb_i,\Hb_i}\left\{\|\Hb_i\|_F : q_i(\xb_k) = F_i(\xb_k) \quad \forall \xb_k\in \iset \right\},
\end{equation}
where $\|\,\|_F$ is the Frobenius norm and the interpolation set
$\iset$ contains between $\nx+1$ and $(\nx+1)(\nx+2)/2$ points satisfying
geometric conditions detailed in \cite{[Wil08],(Con09)}.

The quadratic model $q_i$ cannot be expected to approximate $F_i$ at $\xb$ values
far from the points in $\iset$. Hence, we use a trust region framework, whereby
the model $q_i$ is trusted only close to a base-point $\xh$. Given a radius $\Delta>0$,
we let $\cB=\{\xb\in \R^{\nx}: \|\xb-\xh\| \leq \Delta\}$ denote the spherical
neighborhood within which we trust $q_i$. Correspondingly, the interpolation points
in $\iset$ should not be too far away from $\cB$.

Provided that we know the entire vector of observables $\mathbf{F}(\xb_k)$ at each
$\xb_k\in \iset$, we can obtain a set of model parameters $\{(\gb_i,\Hb_i)\}_{i=1}^{n_d}$,
which we trust inside a common region $\cB$ centered about $\xh$. We can thus form a derivative-free
model of the quadratic \Ref{eq:newton},
\begin{multline}
m(\xh+\db) = f(\xh) + \db^T \sum_{i=1}^{n_d} F_i(\xh) \gb_i
\\
+
\frac{1}{2}\db^T\sum_{i=1}^{n_d}\left( \gb_i \gb_i^T +  F_i(\xh) \Hb_i\right) \db.
 \label{eq:newton2}
\end{multline}
Since we trust this model within $\cB$, we expect that a better $\xb$ can be
obtained by solving the trust region subproblem $\min_{\db}\{m(\xh+\db): \xh+\db\in \cB\}$.
This problem minimizes a quadratic with known derivatives over a compact, convex
region and is hence decidedly easier than the original problem. The observables
are then evaluated at the solution to this subproblem so that we obtain
$\mathbf{F}(\xh+\db)$.

An iterative Newton-like procedure is thus obtained. We note that the trust region 
radius $\Delta$ grows and shrinks from one iteration to the next depending 
on the ratio of the actual decrease obtained at the new point versus the decrease 
predicted by the model in \Ref{eq:newton2}. Similarly, our current estimate of the 
solution, $\xh$, is changed only if an adequate decrease of the function was 
obtained or if we achieved a simple decrease in the function value and the geometry 
of the interpolation set $\iset$ gives us confidence. If we did not adequately 
decrease $f$, we must evaluate at an additional $\xb$ value in order to improve the 
geometry of the set $\iset$ in subsequent iterations.


\subsection{Numerical Parameters}
\label{Subsec-numerics}

The evaluation of the function (\ref{eq:chi2}) at point $\xb$ requires 72 HFB
calculations to generate the $s_{i,j}(\xb)$ points for the 72 nuclei $j$ taken in 
the data set. All HFB calculations were performed with the code {\HFBTHO} 
\cite{[Sto05]}. This code solves the Skyrme-HFB equations in the harmonic oscillator 
({\HO}) basis assuming axial and reflection symmetry. In our optimization, we used 
a spherical basis of $N_{\text{shell}}$=20. The oscillator frequency was determined 
for a given nucleus of mass number $A$ according to the formula 
$\hbar\omega_{\text{oscil}} = 1.2\times {41\over A^{1/3}}$\,MeV \cite{[Dob97]}. 
These two choices guarantee good convergence of the HFB energy with respect to the 
basis size, within about 150 keV of the exact value \cite{[Pei08a]}.

Pairing correlations were described by the pairing ED (\ref{vpair}) with different 
pairing strengths for protons and neutrons, $V_0^n \neq V_0^p$. As customary for 
zero-range pairing forces, a cut-off of $E_{cut}$=60\,MeV is used to truncate 
the quasi-particle space \cite{[Dob84]}. In order to avoid pairing collapse, the 
Lipkin-Nogami prescription was systematically applied according to \cite{[Sto03]}. 

Taking into account the 13 parameters of the Skyrme EDF and the 2 additional parameters 
in the pairing channel requires a 15-parameter search. We have made two additional 
simplifications. First, the tensor coupling constants $C_{0}^{J^{2}}$ and $C_{1}^{J^{2}}$ 
were set to 0. This choice was motivated by our requirement to take as a reference point 
the original SLy4 parameterization of \cite{[Cha95]} where these terms were not included. 
Second, preliminary tests indicated that the isovector effective mass was 
poorly constrained by our data set. As a result, the obtained values of $M_{v}^{*}$ were 
clearly nonphysical with regard to the discussion in \cite{[Les06]}. In the final 
run we therefore discarded $M_{v}^{*}$ from the list of free parameters and kept the 
original SLy4 value.

The final optimization was therefore carried on a set of 12 parameters (10 for 
the Skyrme ED plus 2 pairing strengths):
\begin{multline}
\left\{
\rhoc, \enm/A, \knm, \asym, \lsym, M_s^{* -1},
\right. \\
\left.
C_{0}^{\rho\Delta\rho}, C_{1}^{\rho\Delta\rho}, V_0^n, V_0^p, C_{0}^{\rho\nabla J}, C_{1}^{\rho\nabla J}
\right\},
\label{parameter_set}
\end{multline}
with $C_{0}^{J^{2}} = C_{1}^{J^{2}} = 0$ and $M^{* -1}_{v} = 1.249$.

For scaling purposes, the optimization algorithms tested here require the domain of 
variation of the various parameters $\xb$ to be specified. Since, in practice, 
a large subset of $\xb$ represents symmetric and asymmetric nuclear matter 
properties, the range of variation can be easily set up, even if the exact 
values are not known. Table \ref{table:x_values} in Sec. \ref{Subsubsec-parameters} 
lists the scaling intervals adopted in our optimization.

Following the discussion in Sec. \ref{Subsec-experimental}, our objective function
(\ref{eq:chi2}) contains $D_T$= 3 data types: nuclear masses ($i$=1),
proton rms radii ($i$=2), and OES differences ($i$=3). The total number of
data points is $n_{d}$=108 and breaks down into $n_{1}$=72 nuclear masses (28
spherical and 44 deformed), $n_{2}$=28 rms proton radii, and $n_{3}$=8 OES differences
(4 for neutrons and 4 for protons). The values $d_{i,j}$ of the experimental data
points are given in Sec.~\ref{Subsec-experimental}.

The weights $w_{i}$ in the objective function are used to render all 
quantities dimensionless and to allow for a composite $\chi^{2}$ function. The 
weights must be chosen so that all reduced errors are of the
same order of magnitude: they reflect the expected theoretical uncertainty that 
one can assign to a given observable, which is generally larger 
than the corresponding experimental uncertainty for our data set. In the optimization described here, 
we chose $w_{\text{mass}}$=2.0\,MeV, $w_{\text{radii}}$=0.02\,fm and $w_{\text{OES}}$=50\,keV. 


\section{Results}
\label{Sec-results}

This section contains the optimization results. In Sec. \ref{Subsec-optimized}, 
various properties of the resulting ED parameterizations  are explored. Section 
\ref{Subsec-sensitivity} illustrates the versatility of our approach by providing 
a detailed correlation and sensitivity analysis.


\subsection{Optimized Functionals {\UNEDFNB} and {\UNEDFPRE}: Properties and Stability}
\label{Subsec-optimized}

We give in this section the final parameterization of the Skyrme functionals
{\UNEDFNB} and {\UNEDFPRE} that minimizes the $\chi^{2}$ objective function (\ref{eq:chi2}), 
and we perform a number of checks to probe the quality of the resulting functionals. 
In particular, we test their stability with the RPA response function, check that 
both the spherical and deformed shell structure are on par with other 
parameterizations, and discuss various global performance indicators.


\subsubsection{Solution to the optimization problem}
\label{Subsubsec-parameters}

The optimization of a nuclear energy functional is a complex problem. The 
objective function is the compound result of many different full HFB calculations, 
each the result of a self-consistent iterative procedure. In principle, 
such a function lends itself naturally to parallelization, although the different 
times of calculation of spherical and deformed configurations requires fine load 
balancing. Overall, the cost of one function evaluation can typically amount to 10 
minutes on a standard computer cluster. With such costly evaluations, the number 
of evaluations required to minimize (\ref{eq:chi2}) can rapidly become an issue.

In addition, we have no prior knowledge of the multidimensional surface 
of the objective function in the parameter space. 
There is no guarantee that the parameters are all independent, and,
as we show later, there are correlations between them, which make the
topography of the surface complex.

\begin{figure}[tp]
\center
\includegraphics[width=\linewidth]{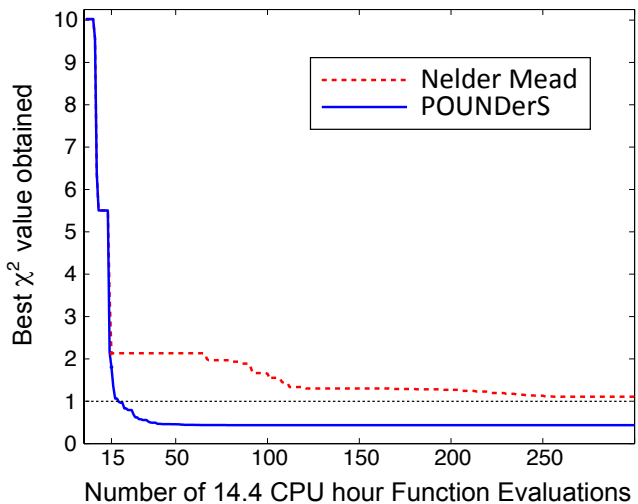}
\caption{(color online) Convergence of the minimization of Eq. (\ref{eq:chi2}) 
with the standard Nelder-Mead algorithm (dashed) and the model-based {\algo} 
(solid line).}
\label{Fig-convergence}
\end{figure}

These observations suggest that two important features of a good optimization
algorithm should be the speed of convergence and the ability to converge to a
true minimum, if only a local one, without being misled by narrow valleys 
and saddle points. Figure \ref{Fig-convergence} shows the performance of the
standard Nelder-Mead (N-M) algorithm, as implemented in the TAO code \cite{[TAO]},
on our objective function, compared with the new model-based algorithm presented 
in Sec. \ref{Subsec-model}. We note that the {\algo} method attains 
a value of $\chi^2$ close to the final one after only 25-30 iterations, whereas 
after more than 300 iterations the N-M algorithm yields a solution that is 
still a factor of 2 away. Moreover, there seems to be a stagnation of the N-M 
method at around 15-65 iterations, which may prematurely suggest that the minimum 
has been found. Yet, in this plateau the $\chi^{2}$  is still about 5 times 
larger than at the final solution.

Table \ref{table:x_values} shows the values of the optimization parameters
(\ref{parameter_set}) at the solution (dubbed 
{\UNEDFNB} in the following). The starting values were given by the
SLy4 parameterization. The most notable change affects the effective mass: 
starting from $M_{s}^{*} \approx 0.7$, the final value is close to 1, which 
ensures a level density more compatible with the empirical one (even though 
there is no obvious reason for this to happen, given the data set employed). 
As will be discussed in Sec.~\ref{Subsec-pairing}, without being steered, the 
optimization gives the correct hierarchy of pairing strengths, namely, 
$|V_0^p| > |V_0^n|$, to reflect the missing momentum-dependence 
and Coulomb contribution, as pointed out in \cite{[Ber09a],[Ang01a],[Les08]}.

\begin{table}[tp]
\begin{center}
\caption{Values $\hat{\xb}$ of the optimization parameters $\xb$ of 
Eq. (\ref{parameter_set}) at the solution with no bounds imposed (Skyrme functional
{\UNEDFNB}). $ \rhoc $ is in fm$^{-3}$; $ \enm/A $, $ \knm $,
$ \asym $, and  $ \lsym $ are in MeV; $ 1/M_{s}^{*} $  is dimensionless; $ C_{t}^{\rho\Delta\rho} $ and $ C_{t}^{\rho\nabla J}$ in MeV\,fm$^5$; and 
$V_0^n$ and $ V_0^p $ 
in  MeV\,fm$^3$.
The range of variation provided 
to the optimization is shown in the column ``Scaling Interval,'' the initial values 
in column $\hat{\xb}^{\rm (init.)}$, and the final values in $\hat{\xb}^{\rm (fin.)}$.
         }
\label{table:x_values}
\begin{ruledtabular}
\begin{tabular}{ll|crr}
$k$ & ${\xb}$  & Scaling Interval & $\hat{\xb}^{\rm (init.)}$ & $\hat{\xb}^{\rm (fin.)}$ \\
\hline
 1. & $ \rhoc $                  & [  +0.14 ,  +0.18] &   +0.160 & 0.151046 \\
 2. & $ \enm/A $                 & [ -17.00,  -15.00] &  -15.972 & -16.0632 \\
 3. & $ \knm $                   & [+170.00, +270.00] & +229.901 & 337.878 \\
 4. & $ \asym $                  & [ +27.00,  +37.00] &  +32.004 & 32.455 \\
 5. & $ \lsym $                  & [ +30.00,  +70.00] &  +45.962 & 70.2185 \\
 6. & $ 1/M_{s}^{*} $            & [  +0.80,   +2.00] &   +1.439 & 0.95728 \\
 7. & $ C_{0}^{\rho\Delta\rho} $ & [-100.00,  -40.00] &  -76.996 & -49.5135 \\
 8. & $ C_{1}^{\rho\Delta\rho} $ & [-100.00, +100.00] &  +15.657 & 33.5289 \\
 9. & $ V_0^n $                  & [-350.00, -150.00] & -258.200 & -176.796 \\
10. & $ V_0^p $                  & [-350.00, -150.00] & -258.200 & -203.255 \\
11. & $ C_{0}^{\rho\nabla J}$    & [-120.00,  -50.00] &  -92.250 & -78.4564 \\
12. & $ C_{1}^{\rho\nabla J}$    & [-100.00,  +50.00] &  -30.750 & 63.9931 \\
\end{tabular}
\end{ruledtabular}
\end{center}
\end{table}

A standard measure of the quality of the optimization is the rms deviation (RMSD) of data type $i$ at the solution $\hat{\xb}$:
\begin{equation}
\text{RMSD}(i) = \sqrt{
\frac{1}{n_i} \sum \limits_{j=1}^{n_{i}}
\left(s_{i,j}(\hat{\xb})-d_{i,j} \right)^2 }.
\end{equation} 
For our set of fit observables, the RMSDs for  various types of data are 
RMSD(mass)=0.966\,MeV,
RMSD(radii)=0.014\,fm, and RMSD(OES)=57\,keV.
For comparison, the value of RMSD(mass)
for SLy4 on the same data set is 9.95 MeV. 

A close examination of Table~\ref{table:x_values} shows that, while most 
of the parameters of {\UNEDFNB} have values in the normally accepted range, the 
incompressibility $\knm$=338\,MeV  is far too large. This would seriously limit the 
usability of {\UNEDFNB} in nuclear structure calculations, in particular in studies of  collective modes such as monopole vibrations.

We therefore performed another minimization, using the same scaling intervals, but imposing hard bounds 
on the NMPs. A similar strategy was adopted in \cite{[Gor06]}, where hard bounds on $\knm$ were imposed during optimization of BSk13 EDF.

\begin{table}[tp]
\begin{center}
\caption{Same as Table \ref{table:x_values} but for the case with bounds
 (Skyrme functional
{\UNEDFPRE}).
}
\label{table:x_valuesb}
\begin{ruledtabular}
\begin{tabular}{ll|crr}
$k$ & ${\xb}$  & Bounds & $\hat{\xb}^{\rm (init.)}$ & $\hat{\xb}^{\rm (fin.)}$ \\
\hline
 1. & $ \rhoc $                  & [+0.15,+0.17] &  +0.160  & 0.160526 \\
 2. & $ \enm/A $                 & [-16.2,-15.8] &  -15.972 & -16.0559  \\
 3. & $ \knm $                   & [+190, +230]  & +229.901 & 230 \\
 4. & $ \asym $                  & [  +28,  +36] &  +32.004 & 30.5429 \\
 5. & $ \lsym $                  & [  +40, +100] &  +45.962 & 45.0804 \\
 6. & $ 1/M_{s}^{*} $            & [ +0.9, +1.5] &   +1.439 & 0.9 \\
 7. & $ C_{0}^{\rho\Delta\rho} $ & [$-\infty, +\infty$] &  -76.996 & -55.2606 \\
 8. & $ C_{1}^{\rho\Delta\rho} $ & [$-\infty, +\infty$] &  +15.657 & -55.6226 \\
 9. & $ V_0^n $                  & [$-\infty, +\infty$] & -258.200 & -170.374 \\
10. & $ V_0^p $                  & [$-\infty, +\infty$] & -258.200 & -199.202 \\
11. & $ C_{0}^{\rho\nabla J}$    & [$-\infty, +\infty$] &  -92.250 & -79.5308 \\
12. & $ C_{1}^{\rho\nabla J}$    & [$-\infty, +\infty$] &  -30.750 & 45.6302 \\
\end{tabular}
\end{ruledtabular}
\end{center}
\end{table}

Table \ref{table:x_valuesb} displays the parameter values of the Skyrme functional 
{\UNEDFPRE} optimized in such a way. 
At convergence, the nuclear incompressibility and scalar effective 
mass appear at their respective bounds of 230 MeV and 1.11 ($1/M_{s}^{*} = 0.9$); 
that is, these NMPs are actively constrained. The rms deviations obtained for 
{\UNEDFPRE} on our set of fit observables are still respectable: RMSD(mass)=1.455\,MeV,
RMSD(radii)=0.016\,fm, and RMSD(OES)=59\,keV.


\subsubsection{Stability check of {\UNEDFNB} and {\UNEDFPRE}}
\label{Subsubsec-stability}

It is known that some Skyrme ED parameterizations  are prone to finite-size instabilities 
\cite{[Bla76],[Cau80a],[Les06],[Kor10]}. For instance, in the time-even channel, the term 
$C_{1}^{\Delta\rho}\rho_{1}\Delta\rho_{1}$ can lead to divergences of the HFB iterative 
procedure. When searching for new functionals, it is therefore crucial to test comprehensively 
the stability of the functional parameterization. Here, the RPA linear response theory 
\cite{[Fet71]} is the tool of  choice. The full RPA response in infinite matter has been 
derived for Skyrme EDFs \cite{[Gar92b],[Mar06a],[Dav09]}, and applications pertaining to 
the stability of Skyrme functionals have been reported in \cite{[Les06],[Sch10]}.

Without entering into details, a general expression for the RPA response function 
$\Pi(\omega,\mathbf{q})$ in SNM can be written as \cite{[Fet71]}
\begin{eqnarray}
 \Pi(\omega,\mathbf{q}) &=& \frac{4\,\Pi_0(\omega,\mathbf{q})}%
    {D(\omega,\mathbf{q})},
  \label{response}
\end{eqnarray}
where $\omega$ is the excitation energy; $\mathbf{q}$ is the transferred momentum (or wave 
number of the density fluctuation); $\Pi_0(\omega,\mathbf{q})$ is the noninteracting 
response (or Lindhard function); and $D(\omega,\mathbf{q})$ is the dielectric function, 
equal to unity in noninteracting SNM.

The value $\Pi(\omega=0,\mathbf{q})$ corresponds to the static susceptibility of
the system to finite-size perturbations. With the above sign convention, 
$\Pi(\omega=0,\mathbf{q})$ should be positive for all values of $\mathbf{q}$ 
and the density $\rho_0$. A change of sign with either variable corresponds to 
$D(\omega,\mathbf{q})=0$; hence, the occurrence of a pole indicating the 
existence of a zero-energy collective mode. In the isospin channel, the 
short-wavelength (high-$\mathbf{q}$) behavior is driven essentially by 
the combination of coefficients $C^{\rho\rho}_1 - C^{\rho\Delta\rho}_1 q^{2}$ 
\cite{[Gar92b],[Les07]}. The magnitude of the latter correlates well 
with the occurrence of instabilities in calculations of finite nuclei.

\begin{figure}[tp]
\center
  \includegraphics[width=\linewidth]{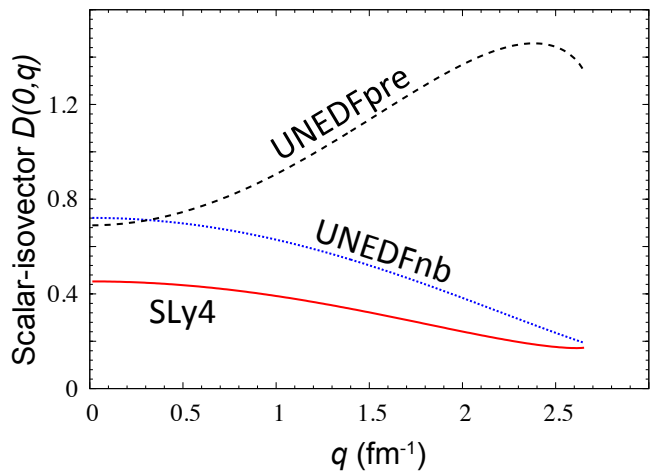}
  \caption{(color online) Dielectric function  $D(\omega=0,\mathbf{q})$ for 
  the scalar-isovector channel in SLy4, {\UNEDFPRE}  and {\UNEDFNB} as a 
  function of the transferred momentum $q$, for $k_F=1.33$\,fm$^{-1}$. }
  \label{fig:rpa}
\end{figure}

Figure \ref{fig:rpa} shows the dielectric function $D(\omega=0,\mathbf{q})$
as a function of $q$, in the scalar-isovector perturbation channel, at
saturation density in SNM. When $D(\omega=0,\mathbf{q})$=0, finite-size
instabilities could potentially develop and hinder the usability of the
functional. This situation does not occur for {\UNEDFNB} and {\UNEDFPRE}, 
which yield a dielectric function even more ``stable'' than SLy4. Varying 
the density, we found that the poles of the response function at $\omega$=0 
occur only for $\rho_{0} \gtrsim 0.22$ fm$^{-3}$. This result does not 
guarantee that other types of instabilities could not develop \cite{[Kor10]}; 
however, we can rule out the most common ones.


\subsubsection{Spherical shell structure}
\label{Subsubsec-shell}

The essence of the nuclear DFT is to be a global theory, whereby one unique
functional (or family thereof) should be used to compute with reasonable
accuracy various properties of atomic nuclei from the lightest to the heaviest.
Many of these properties depend on the single-particle shell structure.
\begin{figure}
\includegraphics[width=\linewidth]{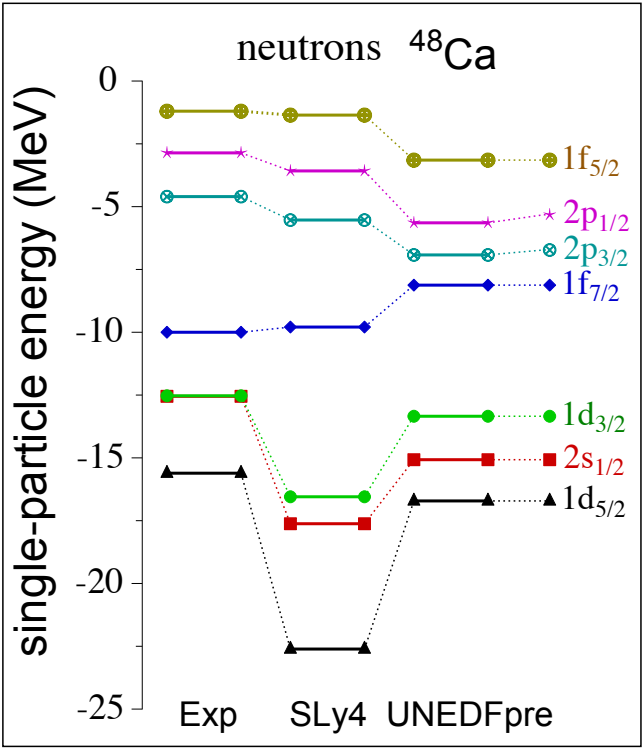}
\caption{(color online) Neutron single-particle energies in $^{48}$Ca  obtained from a HF calculation 
with the functional {\UNEDFPRE}. Experimental s.p. levels \cite{[Sch07]} and SLy4 
results are shown for comparison.}
\label{fig:spCa}
\end{figure}
\begin{figure}
\includegraphics[width=\linewidth]{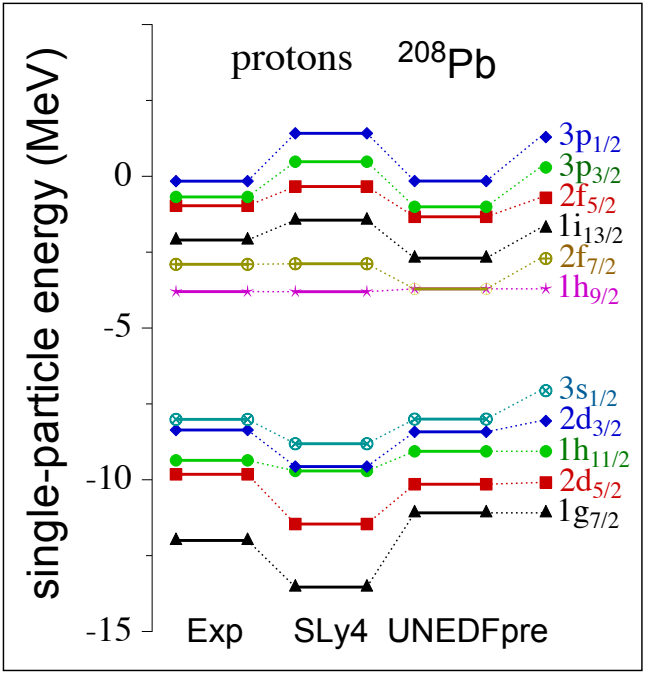}
\caption{(color online) Similar as in Fig.~\ref{fig:spCa} but for 
proton single-particle energies in $^{208}$Pb.}
\label{fig:spPb}
\end{figure}
Figures \ref{fig:spCa} and  \ref{fig:spPb} show, respectively, the neutron s.p. energies in $^{48}$Ca and
proton s.p. energies in $^{208}$Pb obtained with  {\UNEDFPRE}.
They are compared with levels extracted from experiment \cite{[Sch07]} and 
those calculated with   SLy4. In $^{208}$Pb, the overall agreement of 
the proton spectrum is very good. Furthermore, the neutron s.p. levels in $^{208}$Pb 
and proton and neutron levels in $^{132}$Sn (not shown) agree well with experiment. 
As seen in Table \ref{table:x_values}, 
the optimization produces a functional with an effective mass close to 1, 
which is probably the reason the level density in $^{208}$Pb
and $^{132}$Sn  is well reproduced.
Although the overall agreement for s.p.  energies is good, the systematic effect 
of high-$j$ states being slightly too high in energy is seen \cite{[Les07]}.

The neutron single-particle spectrum in $^{48}$Ca is, however, poorly reproduced.
One of the most alarming features is the absence of the magic gap at $N$=28 resulting from a large s.p. level density and a reduced spin-orbit splitting.
The s.p. proton spectrum of $^{48}$Ca is only marginally better with the magic gap at $Z$=20 being  too low, and  the situation is similar in  $^{40}$Ca.

The lack of observables directly probing s.p. properties (such as  spin-orbit splittings or shell-gap sizes) in our objective  function and the bias on heavy nuclei in the set of fit observables  are 
undoubtedly the main  reasons for the poor performance  of {\UNEDFPRE}
regarding the shell structure of light nuclei.
 Nevertheless, one must bear 
in mind that even when the optimization  is exclusively focused  on s.p.
properties, standard  Skyrme functionals perform poorly \cite{[Kor08]}.


\subsubsection{Deformation properties}
\label{Subsubsec-deformed}

The spherical shell structure determines many features of deformed nuclei. Indeed,
the appearance of deformed  states and shape coexistence effects
can be related to s.p.  levels and their couplings through symmetry-violating moments \cite{[Woo92],[Rei99]}. 
Since the shell structure of light nuclei with {\UNEDFPRE} shows large deviations from 
experiment, it is interesting to test whether the new parameterization 
can nonetheless produce sensible deformation properties for medium-mass 
nuclei. To this end, we performed a series of constrained HFB 
calculations for the sequence of Zr isotopes known to exhibit dramatic shape variations as a function of $N$. While nuclei near magic $^{90}$Zr 
are known to be spherical, neutron-rich Zr isotopes with $A\!\!\geq$100 possess
large prolate ground-state deformations, and $^{96-98}$Zr exhibit a 
complex coexistence pattern \cite{[Rei99],[Ska97]}. On the proton-rich 
side, there is strong experimental evidence for large prolate deformation in $N$=$Z$=40 system  $^{80}$Zr \cite{[Rei99]}.

\begin{figure}[h]
\center
\includegraphics[width=\linewidth]{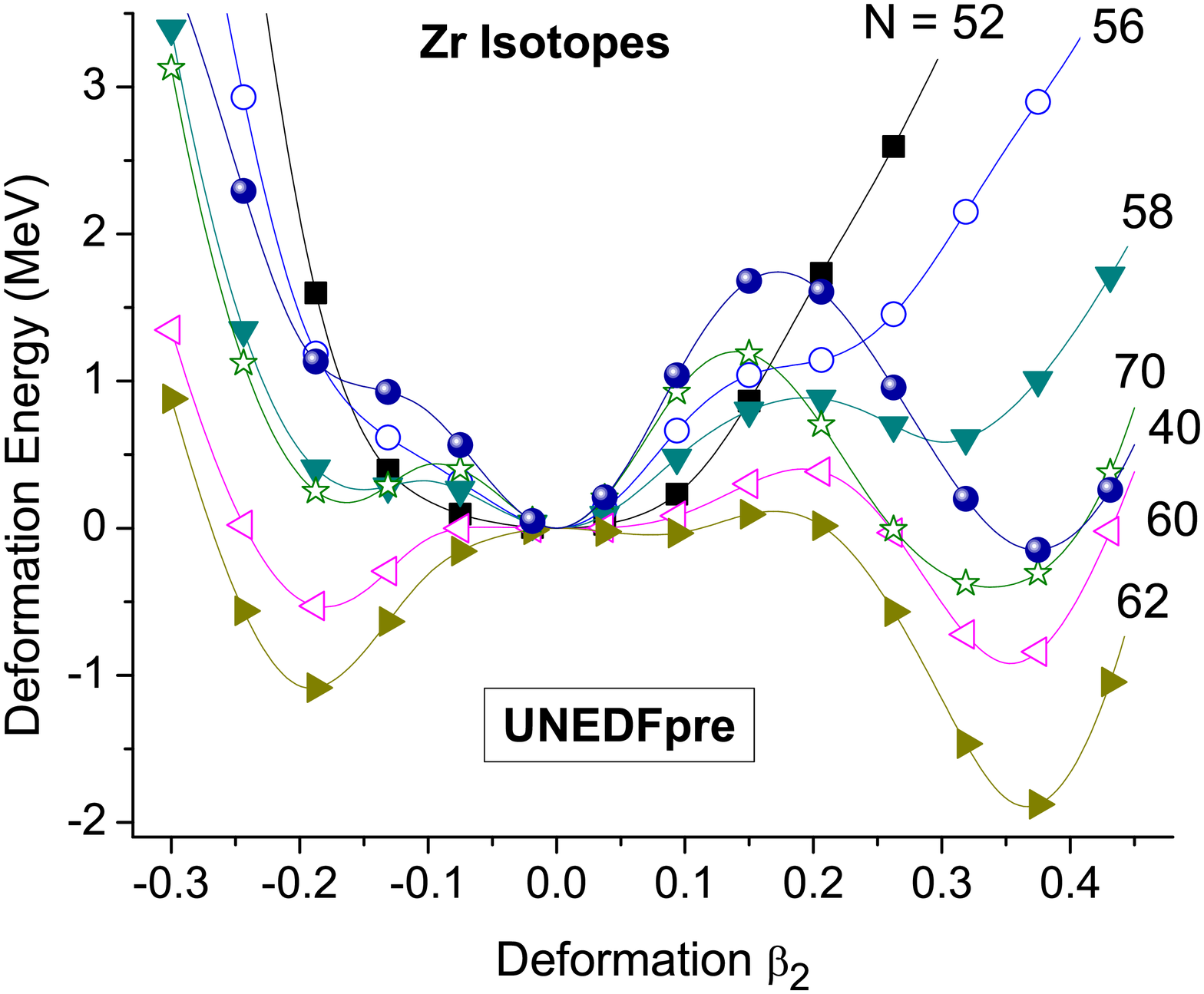}
\caption{(color online) Deformation  energy curves  as functions of the quadrupole deformation 
$\beta_{2}$ for selected even-even Zr isotopes calculated in the HFB+LN approach with {\UNEDFPRE} Skyrme functional.}
\label{fig:def}
\end{figure}

Figure \ref{fig:def} shows the evolution of  HFB+LN deformation energy in the selected even-even Zr isotopes 
as a function of the quadrupole deformation $\beta_{2}$. Each point was computed 
by imposing a constraint on the quadrupole moment 
$\langle \hat{Q}_{2} \rangle \propto \beta_{2}$. Overall, the energy balance 
between spherical and deformed configurations is consistent with experiment. In 
particular, Zr isotopes with $N>58$ are correctly predicted to have prolate 
ground states coexisting with a secondary oblate minimum. 
Also, it is encouraging to see that the ground state of $^{80}$Zr is predicted to be prolate, a feature that is not present in many Skyrme parameterizations \cite{[Rei99]}.


\subsubsection{Global mass table}
\label{mass_table_results}

A good test of any EDF parameterization is its ability to reproduce masses 
across the nuclear chart. Since our objective function contains the binding 
energies of a large  set of nuclei, we expect good agreement with experimental 
data, especially for heavy deformed systems.

\begin{figure}[htp]
  \includegraphics[width=\linewidth]{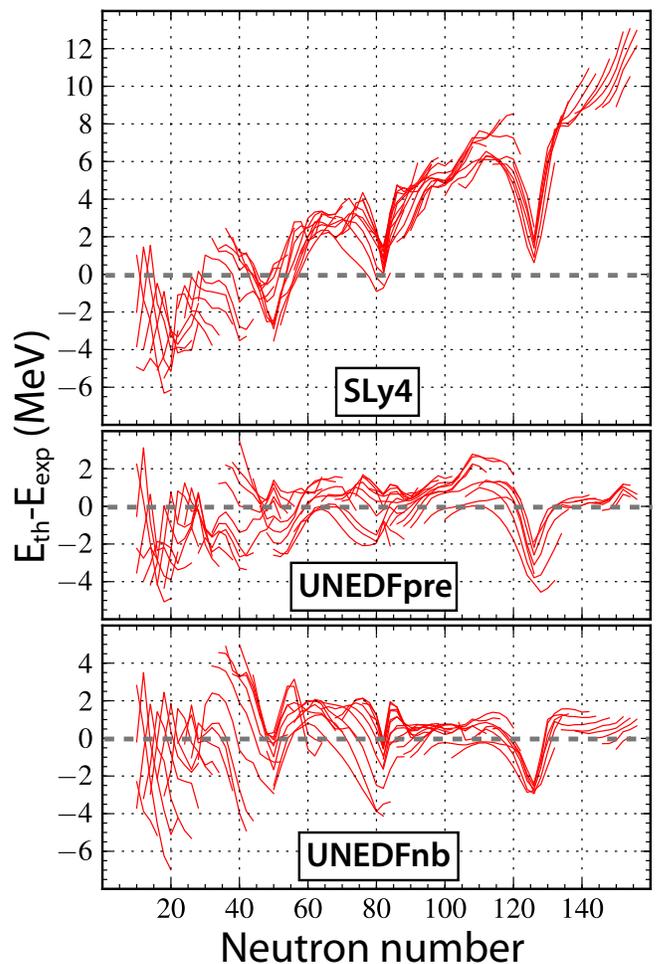}
  \caption{(color online) Binding energy residuals between theory and experiment 
  for 520 even-even nuclei. The HFB+LN results with SLy4 (top) are compared with 
  those of {\UNEDFPRE} (middle) and {\UNEDFNB} (bottom).}
\label{figBEeveneven}
\end{figure}
All even-even nuclei with $N,Z>8$ have been calculated with our two parameterizations
according to the method presented in \cite{[Sto09a]}. Results have been posted for 
visualization and comparison with other EDF parameterizations at 
\url{http://massexplorer.org}. The difference between experimental and theoretical 
binding energies for the 520  even-even nuclei is shown in Fig.~\ref{figBEeveneven}. 
An arclike trend \cite{[DSN04]} is seen for the SLy4 EDF; it has been 
attributed to an overemphasis on doubly magic nuclei during  optimization. By contrast, 
both  {\UNEDFPRE} and {\UNEDFNB} show a much flatter 
behavior, while simultaneously reducing the mass residuals:   
RMSD(mass)=4.80\,MeV for SLy4, and 1.45 MeV and 1.61 MeV for {\UNEDFPRE} and {\UNEDFNB}, 
respectively.

To put things in perspective, we note that the best overall agreement with experimental masses obtained with the Skyrme EDF (on a larger data set that includes light and odd nuclei) is currently 0.582 MeV \cite{[Gor09]}. However, this excellent result was obtained at a price of several corrections on top of the EDF itself. 
In fact, a linear least-squares refit of the standard Skyrme EDF (also using
SLy4 as a starting point) to all even-even nuclear masses achieves a RMSD
of around 1.7 MeV \cite{[Ber05]}. Note also, that the RMSD for the masses
of the {\UNEDFNB} is higher by a 0.16 MeV than for the {\UNEDFPRE} despite the
larger domain available for parameter variation, which is due to the
restricted set of masses used in this work. These figures suggest that
{\UNEDFPRE} is probably within a few hundreds of keV of a globally optimal mass
fit within the parameter space employed here.

\begin{figure}[tp]
\center
  \includegraphics[width=\linewidth]{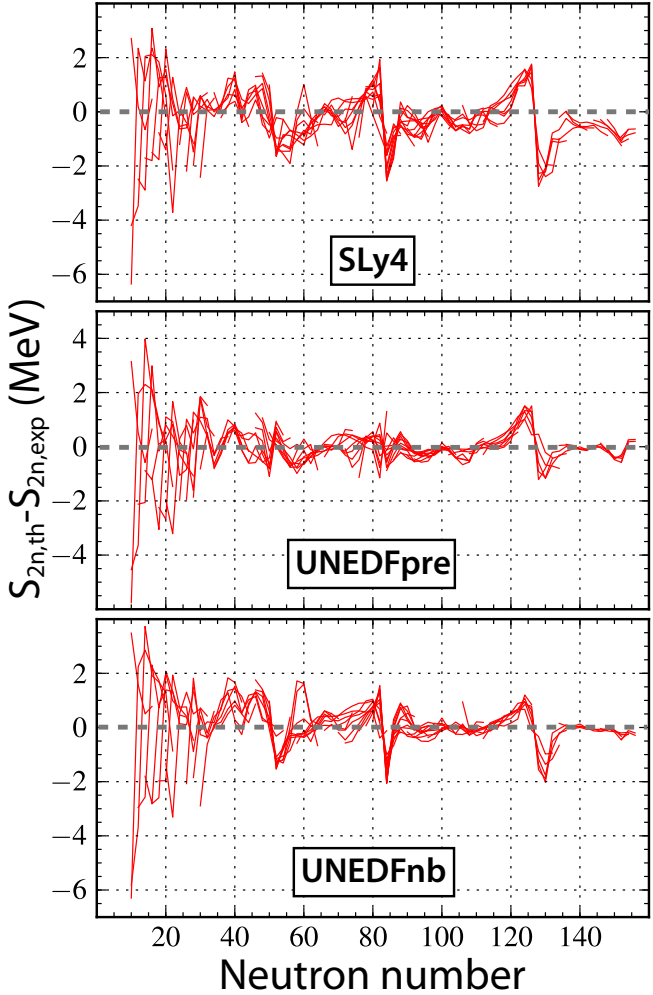}
  \caption{(color online) Two-neutron separation energy residuals between theory and 
  experiment for 520 even-even nuclei. The HFB+LN results with SLy4 (top) are compared 
  with those of {\UNEDFPRE} (middle) and {\UNEDFNB} (bottom).}
\label{figS2neveneven}
\end{figure}

Close examination of Fig.~\ref{figBEeveneven} reveals that, while the global 
trend of binding energy errors has been improved, significant variations
around that global trend still remain. To quantify this, we plot in 
Fig.~\ref{figS2neveneven}  two-neutron separation energy residuals as a 
function of neutron number $N$ for the 520 nuclei of the previous set. The values of  RMSD($S_{2n}$)
for SLy4 and {\UNEDFPRE} are, respectively, 0.99 and 0.76\,MeV, which indicate a  significant 
improvement. If the set of 520 nuclei considered is divided into light ($A<80$) and heavy 
($A \geq 80$) subsets, the respective RMSD($S_{2n}$) values for SLy4 and {\UNEDFPRE} are 
1.41 and 1.45\,MeV for light nuclei, and 0.85 and 0.45\,MeV for heavy nuclei. This result stems from the bias toward heavy nuclei in our  
data set.


\subsubsection{Constraints on pairing strength from optimization}
\label{Subsec-pairing}

Adjusting pairing interaction strengths represents a situation in which, 
by sequentially releasing a constraint on the EDF, one can dramatically 
improve the agreement with a subset of fit observables. The case in point is the interplay 
between pairing and shell structure. Since the shell correction to the binding 
energy favors low s.p. level density around the Fermi level, and the opposite is 
true for pairing contributions, an anticorrelation between these two effects exists 
that results in a cancellation between shell and pairing energies \cite{[Naz94a]}. 
If only total binding energies are subject to optimization, a reasonable fit can 
be obtained by, for example, an unphysical increase in pairing and a simultaneous unphysical 
variation of s.p. shell structure. Indeed, since no data in our 
experimental data set directly probe s.p. energies, the lack of constraints on shell 
structure can dramatically impact pairing properties.

\begin{figure}
\includegraphics[width=0.9\columnwidth]{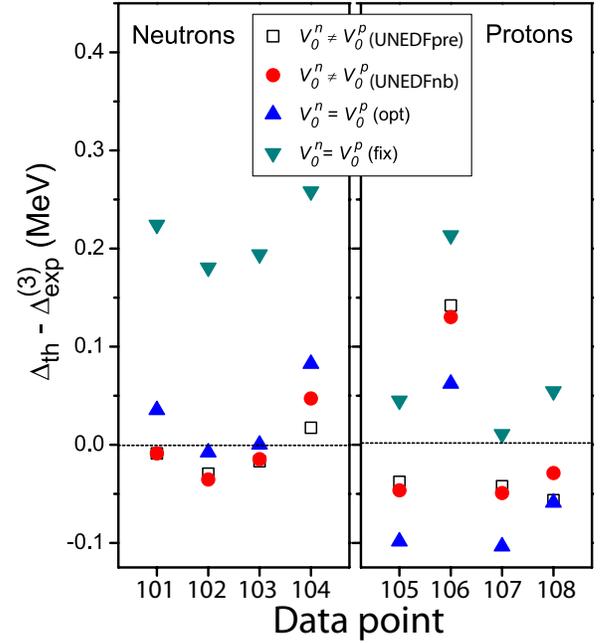}
\caption{(color online) Neutron (left) and proton (right) OES residuals
$\Delta_{th} - \Delta^{(3)}_{exp}$ for the nuclei listed in Table~\ref{table:OES}. 
The results with fixed (non-optimized) values of $V_0^n=V_0^p$ are marked by upside-down triangles. 
The optimized results are marked by triangles ($V_0^n=V_0^p$), dots
($V_0^n \ne V_0^p$; {\UNEDFNB}), and squares ($V_0^n \ne V_0^p$; {\UNEDFPRE}).}
\label{fig-OES}
\end{figure}

Figure~\ref{fig-OES} displays the OES residuals for three variants of calculations. 
In the first variant, the proton and neutron pairing strengths were kept equal and 
fixed at the standard value for SLy4 that yields an average neutron pairing gap in 
$^{120}$Sn equal to the experimental value of 1.245 MeV \cite{[Dob84],[Dob95c]}. 
In this case, the optimization procedure yields shell structure that resulted in 
 overestimated pairing correlations, and the calculated RMSD
for the OEM  is 172 keV.

In the next step, we assumed proton and neutron pairing strengths to be identical 
$V_0^n = V_0^p = V_{0}$, and the constant $V_0$ was included in the optimization 
set. The improvement on pairing energy was immediate, with the rms error on OES 
dropping down to 67\,keV. However, Fig. \ref{fig-OES} clearly shows that OES for 
protons is almost systematically underestimated. This observation calls for using 
different pairing strengths for neutrons and protons, as was suggested by a recent 
large-scale survey  \cite{[Ber09a]}.

Our final optimization run was therefore carried out by considering independent strengths $V_0^n$ and $V_0^p$ in the fit. The rms error on OES has been further reduced to 57\,keV in {\UNEDFNB} and 59\,keV in {\UNEDFPRE}, and the two pairing strengths turn out to be significantly different; see Tables \ref{table:x_values} and \ref{table:x_valuesb}. Apart from possible global physics arguments, this result indicates that this optimization problem benefits from proton and neutron pairing strength being independent parameters.

We conclude this discussion with a word of warning: strictly speaking, the calculation of the OES requires computation of differences of binding energies. In odd nuclei, time-reversal symmetry is broken, time-odd fields are nonzero, and the ground-state should be computed as the lowest quasi-particle excitation of a fully paired vacuum (blocking). Since the correct blocked state is not known beforehand, such calculations are much more involved than in even-even nuclei \cite{[Sch10]}. For this work, where the focus is on the optimization of the Skyrme functional itself and the pairing functional is limited, the extra cost of the proper treatment of odd nuclei was not deemed worth pursuing. 


\subsection{Statistical Analysis of Optimization Results}
\label{Subsec-sensitivity}

From a statistical viewpoint, our optimization problem is also a nonlinear
regression problem. For the true (but unknown) parameter value $\xb_*$ we define
the errors between the theoretical value for the observable of type $i$ in the 
nucleus $j$ and its experimental counterpart as
\begin{equation}
 \eps_{i,j} = \frac{ s_{i,j}(\xb_*)-d_{i,j} } { w_i }.
\label{eq:errors}
\end{equation}
We assume every error $\eps_{i,j}$ is a random variable with expectation 0 and
that all $\eps_{i,j}$ are independent and follow the same distribution. The
optimization presented in Sec. \ref{Subsec-model} estimates $\xb_*$ by the
least-squares estimator
\begin{equation}
\xh = \arg \min_{\xb} \left\{ f(\xb) = \frac{1}{2} \|\Fb(\xb)\|^2\right \}.
\label{eq:lest}
\end{equation}
In the statistical setting, however, the random errors $\epsb = \{ \eps_{i,j} \}$
prevent the random variable $\xh$ from always equaling $\xb_*$.


\subsubsection{Confidence intervals}

To characterize how the parameters change in a neighborhood of $\xb_*$ and
$\xh$, we consider approximate confidence intervals. A $1-\alpha$ 
\emph{confidence interval} $\Omega_k\subset \R$ is one in which we expect the true 
value $x_{k,*}$ to lie $100(1-\alpha)\%$ of the time, that is, with probability 
$P(x_{k,*}\in \Omega_k) = 1-\alpha$.

We note that the assumption of normally distributed residuals, 
$\epsb \sim N(\mathbf{0}, \sigma^2_* I_{\nd})$, made below, is the
strongest one of this regression analysis. As pointed out in \cite{[Toi08]}, 
theoretical (systematic)  errors coming from an imperfect model are
neither random nor generally independent, and their distribution is not
rigorously normal. We carry out a standard analysis nonetheless in order to
investigate constraints applied on our model. Therefore, the confidence intervals given
here are to be understood as ranges of acceptable values for building
parameterizations of this particular model.

Given normally distributed residuals and appropriate regularity conditions
(as in \cite{(NLR89)}, pages 23-25), a $1-\alpha$ confidence interval (CI) centered 
about $\hat{x}_k$ is
\begin{equation}
 \left\{ x_k\in \R:  |x_k-\hat{x}_k| \leq \sqrt{\cov(\xh)_{k,k}} \, \, t_{\nd-\nx,1-\frac{\alpha}{2}} \right\},
\label{eq:ci}
\end{equation}
where $ t_{\nd-\nx,1-\frac{\alpha}{2}}$ is the $1-\frac{\alpha}{2}$
quantile of the t-distribution \cite{[Mon03]} with $\nd-\nx$ degrees
of freedom, and the covariance matrix
$\cov(\xh)=\text{E}[(\xh-\text{E}\xh)(\xh-\text{E}\xh)^T]$.

Using the same notation as in \Ref{eq:newton2}, we use a first-order
approximation of the covariance matrix
\begin{equation}
\hat{\mathbf{V}} \equiv \chi^2(\xh)\left(  \sum_{i=1}^{\nd} \gb_i \gb_i^T\right)^{-1} \approx \cov(\xh),
\label{eq:covariance}
\end{equation}
where parameters $\{\gb_i\}_{i=1,\dots,\nd}$ are found by calculating
central differences on the $2n_x$ points $\left\{\hat{\xb}\pm \eta_k e_k
\right\}_{k=1,\dots,\nx}$, where $\eta_k>0$ is chosen to be small. Although other
approximations to the covariance matrix are possible, the authors of
\cite{[Don87]} state that $\hat{\mathbf{V}}$ is their preferred approximation
because it is ``simpler, less expensive, and more numerically stable'' than
alternative choices. 

\begin{table}[ht]
\begin{center}
\caption{Optimal parameter values of {\UNEDFNB} (no bounds), 95\% confidence 
intervals, percentage of the initial guess for the scaling interval and standard 
deviation $\sigma$.}
\begin{ruledtabular}
\begin{tabular}{ll|cccc}
$k$ & Par. & $ \hat{\xb} $ & 95\% CI & \% of Int. & $\sigma$\\
\hline
 1.& $\rhoc$                  & 0.151046 & [0.149,0.153] & 10 & 0.001 \\
 2.& $\enm/A $                & -16.0632 & [-16.114,-16.013] & 5 & 0.039 \\
 3.& $\knm $                  & 337.878  & [302.692,373.064] & 70 & 26.842 \\
 4.& $\asym $                 &  32.455  & [28.839,36.071] & 72 & 2.759 \\
 5.& $\lsym $                 & 70.2185  & [11.108,129.329] & 296 & 45.093 \\
 6.& $1/M_{s}^{*} $           & 0.95728  & [0.832,1.083] & 21 & 0.096 \\
 7.& $C_{0}^{\rho\Delta\rho}$ & -49.5135 & [-55.786,-43.241] & 21 & 4.785 \\
 8.& $C_{1}^{\rho\Delta\rho}$ & 33.5289  & [-2.246,69.304] & 36 & 27.292 \\
 9.& $V_0^n $                 & -176.796 & [-194.686,-158.906] & 18 & 13.648 \\
10.& $V_0^p $                 & -203.255 & [-217.477,-189.033] & 14 & 10.850 \\
11.& $C_{0}^{\rho\nabla J}$   & -78.4564 & [-85.137,-71.775] & 19 & 5.097 \\
12.& $C_{1}^{\rho\nabla J}$   & 63.9931  & [23.460,104.526] & 54 & 30.921 \\
\end{tabular}
\label{table:CI95}
\end{ruledtabular}
\end{center}
\end{table}

\begin{table}[ht]
\begin{center}
\caption{The same as Table \ref{table:CI95}, except for the {\UNEDFPRE}.}
\begin{ruledtabular}
\begin{tabular}{ll|cccc}
$k$ & Par. & $ \hat{\xb} $ & 95\% CI & \% of Int. & $\sigma$\\
\hline
 1.& $\rhoc$                  & 0.160526 &  [0.160,0.161] & 10 & 0.001  \\
 2.& $\enm/A $                & -16.0559 &  [-16.146,-15.965] & 45 & 0.055 \\
 3.& $\knm $                  &     230  &  -- & -- & --\\
 4.& $\asym $                 & 30.5429  &  [25.513,35.573] & 126 & 3.058 \\
 5.& $\lsym $                 & 45.0804  &  [-20.766,110.927] & 219 & 40.037 \\
 6.& $1/M_{s}^{*} $           &     0.9  &  -- & -- & -- \\
 7.& $C_{0}^{\rho\Delta\rho}$ & -55.2606 &  [-58.051,-52.470] & 9 & 1.697 \\
 8.& $C_{1}^{\rho\Delta\rho}$ & -55.6226 &  [-149.309,38.064] & 94 & 56.965 \\
 9.& $V_0^n $                 & -170.374 &  [-173.836,-166.913] & 3 & 2.105 \\
10.& $V_0^p $                 & -199.202 &  [-204.713,-193.692] & 6 & 3.351 \\
11.& $C_{0}^{\rho\nabla J}$   & -79.5308 &  [-85.160,-73.901] & 16 & 3.423 \\
12.& $C_{1}^{\rho\nabla J}$   & 45.6302  &  [-2.821,94.081] & 65 & 29.460 \\
\end{tabular}
\label{table:CI95b}
\end{ruledtabular}
\end{center}
\end{table}

Table~\ref{table:CI95} shows the 95\% confidence intervals, and standard 
deviations obtained when $\eta_k$ is chosen to be $10^{-5}$ times the size 
of the scaling interval of parameter $x_k$. Standard deviations $\sigma$ 
are square roots of the diagonal components of the covariance matrix 
$\cov(\xh)$ and are often also referred to as errors of parameters. 

Confidence intervals can therefore be valuable for testing the completeness 
of a given data set. In our case no data on giant resonances were included, 
which may explain why $\asym$ and $\lsym$ remain imprecise. Similarly, our 
data set does not contain sufficiently many neutron-rich nuclei and/or entire 
isotopic sequences to pin down the isovector coupling constants. We also 
remark that the analysis based on confidence intervals is straightforward to 
perform once the (computationally-intensive) covariance matrix is known.


\subsubsection{Sensitivity Analysis}

\begin{table*}[!]
\caption{\label{tab:cor} Correlation matrix (\ref{eq:correlation}) 
for the {\UNEDFNB} parameter set (no bounds). }
\begin{center}
\begin{ruledtabular}
\begin{tabular}{c|cccccccccccc}
 $ \rhoc $                & 1.00 & & & & & & & & & & &\\
 $ \enm/A $               & -0.04 & 1.00 & & & & & & & & & &\\
 $ \knm $                 & -0.87 & 0.16 & 1.00 & & & & & & & & &\\
 $ \asym $                & -0.05 & -0.72 & -0.29 & 1.00 & & & & & & & &\\
 $ \lsym $                & -0.09 & -0.62 & -0.23 & 0.97 & 1.00 & & & & & & &\\
 $ 1/M_{s}^{*} $          & -0.05 & 0.05 & 0.09 & -0.10 & -0.10 & 1.00 & & & & & &\\
 $C_{0}^{\rho\Delta\rho}$ & -0.23 & 0.24 & 0.34 & -0.25 & -0.20 & -0.86 & 1.00 & & & & &\\
 $C_{1}^{\rho\Delta\rho}$ & -0.22 & 0.29 & 0.34 & -0.65 & -0.76 & -0.08 & 0.28 & 1.00 & & & &\\
 $ V_0^n $                & 0.02 & -0.02 & -0.06 & 0.06 & 0.06 & -0.99 & 0.87 & 0.12 & 1.00 & & &\\
 $ V_0^p $                & 0.01 & -0.14 & -0.10 & 0.26 & 0.27 & -0.95 & 0.78 & -0.07 & 0.93 & 1.00 & &\\
 $ C_{0}^{\rho\nabla J}$  & 0.07 & -0.03 & 0.04 & -0.14 & -0.17 & -0.72 & 0.78 & 0.32 & 0.73 & 0.65 & 1.00 &\\
 $ C_{1}^{\rho\nabla J}$  & -0.07 & -0.35 & -0.12 & 0.58 & 0.66 & 0.06 & -0.26 & -0.64 & -0.08 & 0.05 & -0.38 & 1.00\\
\hline
& $\rhoc$ & $\enm/A$ & $\knm$ & $\asym$ & $\lsym$ &
$1/M_{s}^{*}$ & $C_{0}^{\rho\Delta\rho}$ & $C_{1}^{\rho\Delta\rho}$ &
$V_0^n$ & $V_0^p$ & $ C_{0}^{\rho\nabla J}$ & 
$C_{1}^{\rho\nabla J}$ \\
\end{tabular}
\end{ruledtabular}
\end{center}
\end{table*}

\begin{table*}[!]
\caption{\label{tab:corb} Correlation matrix (\ref{eq:correlation}) 
for the {\UNEDFPRE} parameter set. }
\begin{center}
\begin{ruledtabular}
\begin{tabular}{c|cccccccccccc}
 $ \rhoc $                & 1.00 & & & & & & & & & & &\\
 $ \enm/A $               & -0.28 & 1.00 & & & & & & & & & &\\
 $ \knm $                 & --    & --    & -- & & & & & & & & &\\
 $ \asym $                & -0.10 & -0.88 & -- & 1.00 & & & & & & & &\\
 $ \lsym $                & -0.17 & -0.80 & -- & 0.97 & 1.00 & & & & & & &\\
 $ 1/M_{s}^{*} $          & --    & --    & -- & -- & -- & -- & & & & & &\\
 $C_{0}^{\rho\Delta\rho}$ &  0.09 & 0.80  & -- & -0.81 & -0.74 & -- &  1.00 & & & & &\\
 $C_{1}^{\rho\Delta\rho}$ &  0.20 & 0.35  & -- & -0.47 & -0.66 & -- &  0.23 &  1.00 & & & &\\
 $ V_0^n $                &  0.02 & 0.21  & -- & -0.23 & -0.25 & -- &  0.23 &  0.23 &  1.00 & & &\\
 $ V_0^p $                & -0.13 & -0.42 & -- &  0.52 &  0.56 & -- & -0.29 & -0.45 & -0.14 & 1.00 & &\\
 $ C_{0}^{\rho\nabla J}$  &  0.37 & -0.14 & -- &  0.02 & -0.00 & -- &  0.44 & -0.02 &  0.09 & 0.16 &  1.00 &\\
 $ C_{1}^{\rho\nabla J}$  & -0.06 & -0.18 & -- &  0.27 &  0.33 & -- & -0.38 & -0.20 & -0.01 & 0.00 & -0.37 & 1.00\\
\hline
& $\rhoc$ & $\enm/A$ & $\knm$ & $\asym$ & $\lsym$ &
$1/M_{s}^{*}$ & $C_{0}^{\rho\Delta\rho}$ & $C_{1}^{\rho\Delta\rho}$ &
$V_0^n$ & $V_0^p$ & $ C_{0}^{\rho\nabla J}$ & 
$C_{1}^{\rho\nabla J}$ \\
\end{tabular}
\end{ruledtabular}
\end{center}
\end{table*}

The covariance matrix depends on the scaling of the parameters; hence, we will
work with the standard correlation coefficient,
\begin{equation}
R_{k,l}=\frac{\cov(x_k,x_l)}{\sqrt{\var(x_k)\var(x_l)}}, 
\label{eq:correlation}
\end{equation}
which captures the (positive or negative) correlation between parameters $x_k$ 
and $x_l$. Tables~\ref{tab:cor}-\ref{tab:corb} provide the approximate $n_{x}\times n_{x}$ 
correlation matrix $\mathbf{R}$ calculated when $\eta_k$ is chosen to be $10^{-5}$ 
the size of the interval of interest of parameter $x_k$, for the solutions {\UNEDFPRE} 
and {\UNEDFNB}, respectively.

Overall, Tables \ref{tab:cor}-\ref{tab:corb} show that most parameters are 
interdependent, although the number of significant correlations with $|R_{kk}|\geq 0.8$ 
is  small. For {\UNEDFNB}, where all parameters are free, we note that two 
pairs of NM parameters are well correlated: $\knm$ is 87\% correlated with $\rhoc$ 
\cite{[Coc04a],[Che09]}, while $\asym$ is 97\% correlated with $\lsym$ \cite{[Ton84],[Rei06],[War09]}. The 
value of the (inverse of the) effective mass appears well correlated with
both pairing strengths. We also notice strong correlation between the pairing strengths 
and the isoscalar spin-orbit coupling constants. Both observations reflect the 
interplay between single-particle level density  and pairing discussed
in Sec.~\ref{Subsec-pairing}. 
We also notice that the proton pairing strength is significantly 
correlated with the neutron pairing strength.

In the case of the {\UNEDFPRE} parameterization, $\knm$ and $1/M_{s}^{*}$ are removed from the 
sensitivity analysis. Nevertheless, we note that the various correlations between parameters overall 
remain, even if they are attenuated compared to the no-bound case. 

\begin{figure}[h]
\center
\includegraphics[width=\linewidth]{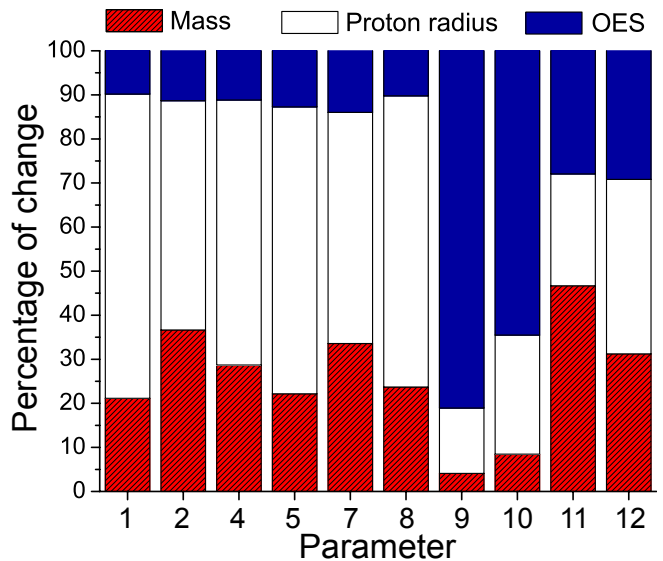}
 \caption{\label{fig:sens_bar} (color online) Sensitivity of the parameters of  
 {\UNEDFPRE}  to different data types entering  $\chi^2$. The EDF parameters are 
 labeled as in Table~\ref{table:CI95}.}
\end{figure}
Next, we illustrate how sensitive the parameters $x_k$ are to the different 
data types entering  $\chi^2$: masses, proton radii, and OES. Here we focus only on the {\UNEDFPRE} 
parameterization, as it is more realistic. We define the 
$n_{x}\times n_{d}$ Jacobian matrix $J(\xb)$ of the residuals as 
$J = ( \gb_{1},\dots,\gb_{n_{d}})$, that is, the matrix formed as the 
juxtaposition of the $n_{d}$ column-vectors $\gb_{i}$ ($n_{d}$ being 
the number of data, $n_{x}$ the number of parameters). The $n_{x}\times n_{d}$ 
sensitivity matrix $S$ is
\begin{equation}
S(\xb) = \left[ J(\xb)J^{T}(\xb) \right]^{-1}J(\xb).
\end{equation}

For each line in the sensitivity matrix (each parameter), we can 
compute the partial sums over each of the three types of data. This computation
gives us a measure of the change of the parameter under a global 
change of all the data of a given type. Figure~\ref{fig:sens_bar} shows 
the relative change of parameter $x_k$ when such an {\it average} 
datum of an observable  is changed. For example, for $i$= 1 (masses), 
it shows the change in $x_k$ under a variation of all experimental 
masses.

All of the bars in Fig.~\ref{fig:sens_bar} have been renormalized to 
unity, and only relative strengths between mass, radii, and OES data 
are shown. A large percentage contribution from data type $i$ means 
that $x_k$ is very sensitive to changes in $i$, and other data types 
have little impact on it at the convergence point. As expected, pairing strengths (parameters 9 and 
10) are primarily affected by OES data. It is worth noting the very similar sensitivity 
of the spin-orbit coupling constants (parameters 11 and 12) on all 3 types of data. Also, nuclear matter parameters appear to be significantly more dependent on the proton radius than other coupling constants. This is not surprising, considering the relation between the saturation density and the Wigner–Seitz radius.

\begin{figure}[tp]
\center
\includegraphics[width=\linewidth]{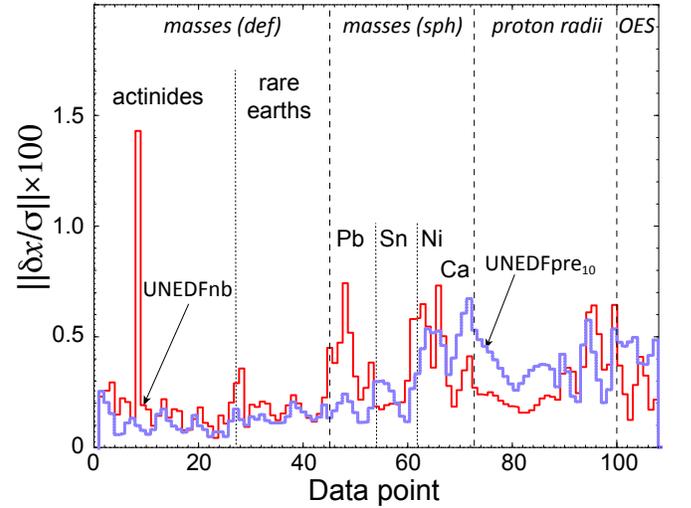}
 \caption{(color online) Overall change in $\xh$ (\ref{eq:change_normed}) for the {\UNEDFNB} 
 and {\UNEDFPRE} parameter sets when data $d_{i,j}$ change by $0.1 w_i$ one-by-one. 
 The labeling of data points is consistent with Tables~\ref{table:edef}-\ref{table:OES}.}
 \label{fig:sens}
\end{figure}

The integrated information contained in Fig.~\ref{fig:sens_bar} cannot assess 
the impact of a particular data piece on model parameters; hence, a more 
detailed analysis is needed. To this end, for each experimental observable 
$d_{i,j}$ (masses, radii, OES), we compute the global change in $\xh$ as 
individual data $d_{i,j}$ change (one-by-one) by $0.1 w_i$, namely, 200 keV 
for masses, 0.002 fm for proton radii, and 50 keV for OES. In this way we can, 
for instance, evaluate the possible importance of some new experimental 
observable on a given model \cite{[Rei10]}. Figure~\ref{fig:sens} shows the 
quantity
\begin{equation}
\| \delta\gras{x}/\sigma \| = \sqrt{\sum_{k=1}^{n_{x}} \left( \frac{\delta x_{k}}{\sigma_{k}} \right)^{2} }
\label{eq:change_normed},
\end{equation}
with
$\delta x_{k}$ being the change in the value of the parameter $x_{k}$ under 
a change of the data $d_{i,j}$, 
for all $\nd$=108 data points. This is nothing but the norm of the total change 
in units of the standard deviation $\sigma_k$, defined as before by 
$\sigma_k = \sqrt{\text{Cov}(x_{k},x_{k})}$. Large changes in $\xh$ mean 
that the parameter values are highly sensitive to the particular 
value of $d_{i,j}$. 

In principle, the sensitivity analysis can be performed at any point $\xb$ of the $n_{x}$-dimensional parameter space: for a given scalar function $f(\xb)$ of the type (\ref{eq:chi2}), the sensitivity at point $\xb$ is only based on the local gradient. In particular, it is totally independent of the procedure that leads to the specific selection of $\xb$. It only depends on the degrees of freedom, i.e., free parameters, retained in  $f(\xb)$. 

The question of degrees of freedom is highly relevant in the context of the {\UNEDFPRE} parameter set, where two parameters are actively constrained at the solution. These constraints have been directly implemented by restricting the domain where the $\chi^{2}$ function is evaluated, and not by modifying the function by adding a penalty.

The net result of imposing the bound constraints is that only 10 parameters out of twelve are allowed to change near the end of the optimization process. One has, therefore, two options as far as the sensitivity analysis is concerned:
(i) Remove these two parameters from the set of active parameters and calculate the Jacobian $J(\xb)$ and the sensitivity matrix $S(\xb)$ with only $n_{x} - 2$=10  parameters;
(ii) Keep all $n_{x}$=12 parameters in the calculation of the Jacobian and doing a tangent plan approximation to obtain the relevant covariance and sensitivity matrices. In this way, the uncertainties of the other 10 parameters are affected by the local fluctuations of the surface induced by these two actively constrained parameters.

The  alternative (i) boils down to computing the gradient at point $\hat{\xb}$ in a 10-parameter subspace of the original 12-parameter space. In this subspace, $\hat{\xb}$ is the stable point  of the $\chi^{2}$ function. The curve labeled {\UNEDFPRE}$_{10}$ in Fig. \ref{fig:sens} corresponds to this approach. This sensitivity response is compared with that performed at the free minimum $\hat{\xb}^{\text{(nb)}}$ of  $\chi^{2}$  in the full 12-parameter space of {\UNEDFNB}. Since both sets correspond to (unconstrained) minima in their respective spaces, the overall changes in $\xh$ are of the same order of magnitude and they are very small, $\| \delta\gras{x}/\sigma \|\approx 0.01$. This indicates that the  set of fit observables has been chosen very consistently. Indeed the
mass of deformed $^{254}$Fm is the single observable that yields the noticeable parameter  variations around the  {\UNEDFNB} minimum while in the case of {\UNEDFPRE}$_{10}$ no sensitivity to a single piece of data can be noticed.

By contrast, within the alternative (ii), $\hat{\xb}$ is not an unconstrained minimizer in the full 12-parameter space. In such an approach, we find the overall sensitivity to be about 2 orders of magnitude larger than what is depicted in Fig. \ref{fig:sens}. This reflects the fact that  $x_{3}$ and $x_{6}$ are far away from the unconstrained minimum in that space, at the same time being strongly correlated with other parameters. In this case, the sensitivities depend strongly on the actual value of $\hat{\xb}$ and the way the domain is constrained. For this reason, the option (ii) is of no practical interest in the comparison with the no-bounds results.


\section{Conclusions}\label{conclusions}

One of the major challenges for the low-energy nuclear theory is to construct
the global nuclear energy density functional of spectroscopic quality, rooted 
in microscopic theory. An important element of this program is to optimize the 
parameters of the functional on a set of experimental observables and selected 
theoretical pseudo-data. This work shows how such an optimization can be done
by using modern optimization algorithms and nonlinear regression analysis.

The purpose of this study was to optimize the standard Skyrme functional based 
on a set of experimental data (masses, charge radii, and odd-even mass differences) 
pertaining to 72 spherical and deformed nuclei amenable to a mean-field description. 
The new model-based optimization algorithm {\algo} was compared with other 
standard derivative-free optimization methods such as Nelder-Mead  and was 
found to be significantly better in terms of speed, accuracy, and precision.

The optimization was carried out at the fully self-consistent, deformed
Hartree-Fock-Bogoliubov level. Here, we took advantage of the efficient DFT solver 
{\HFBTHO} optimized  in the first phase of the project.  We have implemented various improvements that enable us to quickly compute global self-consistent  
mass tables. This capability is essential for optimization. 

As a result of the twelve-parameter optimization of Skyrme EDF, we arrived at two solutions. The first one  corresponds to a 
minimum (stable point) in the considered parameter space. The corresponding functional {\UNEDFNB} describes well the assumed set of fit observables, but its  incompressibility parameter is too large, as this property has not been well constrained by our data set. The second optimization was carried out assuming hard bounds on the nuclear matter parameters. For the bound-constrained 
solution, the nuclear incompressibility and scalar effective mass appear at their respective bounds. The resulting parameter set {\UNEDFPRE} gives good agreement with experimental masses, radii, and deformations and seems to be free of finite-size instabilities. In particular, for two-neutron separation energies and masses of  even-even heavy nuclei with  $A>80$, {\UNEDFPRE} yields the rms deviation of 0.45\,MeV and 1.2\,MeV, respectively, which is a satisfying result. We emphasize that the original Skyrme EDFs seem to be inherently limited in this respect, as demonstrated in \cite{[Ber05]}, unless specific corrections are introduced. Our result is therefore in line with the best expectations one could have for such EDFs. Nevertheless, the lack of specific constraints on the shell structure in our data set  implies that single-particle levels of light nuclei are not well reproduced. For that reason, {\UNEDFPRE} may not yet be recommended for truly  global applications across the chart of the nuclides. However, this functional is expected to work well for heavy nuclei and should  be considered as a reference against which more advanced EDFs will be benchmarked.

We have also applied full-fledged regression diagnostics  on {\UNEDFNB} and 
{\UNEDFPRE}, focusing on statistical correlations between ED parameters and the 
sensitivity of parameters to variations in fit observables.   To this end, we 
computed and analyzed the 
correlation and sensitivity matrices at the optimal parameter set. 
This kind of nonlinear regression  analysis is expected to be helpful when designing next generation EDFs.  
Moreover, the statistical tools presented in this study can be used to pinpoint specific 
nuclear observables that  are expected to strongly affect the developments of 
the nuclear universal density functional.


\begin{acknowledgments}
We thank Peter Kl{\"u}pfel for his help with  the experimental database used in 
this work. This work was supported by the Office of Nuclear Physics, U.S. Department 
of Energy under Contract Nos. DE-FC02-09ER41583 (UNEDF SciDAC Collaboration), 
DE-FG02-96ER40963 and DE-FG02-07ER41529 (University of Tennessee),  DE-FG0587ER40361 (Joint Institute 
for Heavy Ion Research), and DE-AC0Z-06CA11357 (Argonne National Laboratory). 
Computational resources were provided through an INCITE award ``Computational 
Nuclear Structure" by the National Center for Computational Sciences (NCCS) and 
National Institute for Computational Sciences (NICS) at Oak Ridge National Laboratory, and 
through an award by the Laboratory Computing Resource Center at Argonne National Laboratory.
\end{acknowledgments}



\begin{thebibliography}{100}

\bibitem{[IUPAP09]}
{{\it Research Facilities in Nuclear Physics}, IUPAP Report 41.}

\bibitem{riatheory}
{ {\it RIA Theory Bluebook: A Road Map},\\
  http://www.orau.org/ria/RIATG/Blue\_Book\_FINAL.pdf}.

\bibitem{[Top50]}
{Top 500, http://www.top500.org}.

\bibitem{[Ber07]}
{G.F. Bertsch, D.J. Dean, and W. Nazarewicz, SciDAC Review, Winter 2007, p.
  42}.

\bibitem{[Ben03]}
{M. Bender, P.-H. Heenen, and P.-G. Reinhard, Rev. Mod. Phys. {\bf 75}, 121
  (2003)}.

\bibitem{[Sto09]}
{M.V. Stoitsov, J. Mor\'e, W. Nazarewicz, J.C. Pei, J. Sarich, N. Schunck, A.
  Staszczak, and S. Wild, J. Phys., Conference Series {\bf 180}, 012082
  (2009)}.

\bibitem{[Sch10]}
{N. Schunck, J. Dobaczewski, J. Mor\'e, J. McDonnell, W. Nazarewicz, J. Sarich,
  and M. V. Stoitsov, Phys. Rev. C {\bf 81}, 024316 (2010)}.

\bibitem{[Sto06]}
{M.V. Stoitsov, J. Dobaczewski, W. Nazarewicz, and P. Borycki, Int. J. Mass
  Spectrometry {\bf 251}, 243 (2006)}.

\bibitem{[Sto09a]}
{M.V. Stoitsov, W. Nazarewicz, N. Schunck, Int. J. Mod. Phys.  E {\bf 18} 816 (2009).}

\bibitem{[Ber05]}
{G.F. Bertsch, B. Sabbey, and M. Uusn\"akki, Phys. Rev. C {\bf 71}, 054311
  (2005)}.

\bibitem{[Kor08]}
{M. Kortelainen, J. Dobaczewski, K. Mizuyama, and J. Toivanen, Phys. Rev. C
  {\bf 77}, 064307 (2008)}.

\bibitem{[Dru10]}
{J.E. Drut, R.J. Furnstahl, and L. Platter, Prog. Part. Nucl. Phys. {\bf 64},
  120 (2010)}.

\bibitem{[Geb09]}
{B. Gebremariam, T. Duguet, and S.K. Bogner, nucl-th/arXiv:0910.4979 (2009).}

\bibitem{[Geb10]}
{B. Gebremariam, S.K. Bogner, and T. Duguet, nucl-th/arXiv:1003.5210 (2010).}

\bibitem{[Car10]}
{B.G. Carlsson and J. Dobaczewski, nucl-th/arXiv:1003.2543 (2010)}.

\bibitem{[Car08]}
{B.G. Carlsson, J. Dobaczewski, and M. Kortelainen, Phys. Rev. C {\bf 78},
  044326 (2008)}.

\bibitem{[Car09]}
{B.G. Carlsson, J. Dobaczewski, J. Toivanen, and P. Vesely,
  nucl-th/arXiv:0912.3230 (2009).}

\bibitem{[Cha95]}
{E. Chabanat, P. Bonche, P. Haensel, J. Meyer, and R. Schaeffer, Phys.
  Scr. {\bf T56}, 231 (1995)}.

\bibitem{[Dec80]}
{J. Decharg\'e and D. Gogny, Phys. Rev. C {\bf 21}, 1568 (1980)}.

\bibitem{[Ber91]}
{G.F. Bertsch and H. Esbensen, Ann. Phys. (N.Y.) {\bf 209}, 327 (1991)}.

\bibitem{[Toi08]}
{J. Toivanen, J. Dobaczewski, M. Kortelainen, and K. Mizuyama, Phys. Rev. C
  {\bf 78}, 034306 (2008)}.

\bibitem{[Klu09]}
{P. Kl\"upfel, P.-G. Reinhard, T.J. Burvenich, and J.A. Maruhn, Phys. Rev. C
  {\bf 79}, 034310 (2009)}.

\bibitem{[Bar82]}
{J. Bartel, P. Quentin, M. Brack, C. Guet, and H.B. H{\aa}kansson, Nucl. Phys.
  A {\bf 386}, 79 (1982)}.

\bibitem{[Ton00]}
{F. Tondeur, S. Goriely, J.M. Pearson, and M. Onsi, Phys. Rev. C {\bf 62},
  024308 (2000)}.

\bibitem{[Gor07]}
{S. Goriely, M. Samyn, and J.M. Pearson, Phys. Rev. C {\bf 75}, 064312 (2007)}.

\bibitem{[Dug10]}
{T. Duguet and J. Sadoudi, J. Phys. G: Nucl. Part. Phys. {\bf 37} 064009
  (2010).}

\bibitem{[She00a]}
{J.A. Sheikh and P. Ring, Nucl. Phys. A {\bf 665}, 71 (2000)}.

\bibitem{[She02]}
{J.A. Sheikh, P. Ring, E. Lopes, and R. Rossignoli, Phys. Rev. C {\bf 66},
  044318 (2002)}.

\bibitem{[Sto07]}
{M.V. Stoitsov, J. Dobaczewski, R. Kirchner, W. Nazarewicz, and J. Terasaki,
  Phys. Rev. C {\bf 76}, 014308 (2007)}.

\bibitem{[Dob09g]}
{J. Dobaczewski, J. Phys. G: Nucl. Part. Phys. {\bf 36}, 105105 (2009)}.

\bibitem{[Lac09]}
{D. Lacroix, T. Duguet, and M. Bender, Phys. Rev. C {\bf 79}, 044318 (2009)}.

\bibitem{[Dug09]}
{T. Duguet, M. Bender, K. Bennaceur, D. Lacroix, and T. Lesinski, Phys. Rev. C
  {\bf 79}, 044320 (2009)}.

\bibitem{[Ben09]}
{M. Bender, T. Duguet, and D. Lacroix, Phys. Rev. C {\bf 79}, 044319 (2009)}.

\bibitem{[Fri86]}
{J. Friedrich and P-G. Reinhard, Phys. Rev. C {\bf 33}, 335 (1986)}.

\bibitem{[Rei10]}
{P.-G. Reinhard and W. Nazarewicz, Phys. Rev. C, in press (2010);
  nucl-th/arXiv:1002.4140.}

\bibitem{(Rin80)}
P.\ Ring and P.\ Schuck.
\newblock {\em The Nuclear Many-Body Problem}.
\newblock Springer-Verlag, Berlin, 1980.

\bibitem{[Per04]}
{E. Perli\'nska, S.G. Rohozi\'nski, J. Dobaczewski, and W. Nazarewicz, Phys.
  Rev. C {\bf 69}, 014316 (2004)}.

\bibitem{[Eng75]}
{Y.M. Engel, D.M. Brink, K. Goeke, S.J. Krieger, and D. Vautherin, Nucl. Phys.
  A {\bf 249}, 215 (1975)}.

\bibitem{[Dob02c]}
{J. Dobaczewski, W. Nazarewicz, and M. V. Stoitsov, Eur. Phys. J. A {\bf 15},
  21 (2002)}.

\bibitem{[Cha97]}
{E. Chabanat, P. Bonche, P. Haensel, J. Meyer, and R. Schaeffer, Nucl. Phys. A
  {\bf 627}, 710 (1997)}.

\bibitem{[Sto07b]}
{J. R. Stone and P.-G. Reinhard, Prog. Part. and Nucl. Phys. {\bf 58}, 587
  (2007)}.

\bibitem{[Pie09]}
{J. Piekarewicz, J. Phys. G: Nucl. Part. Phys. {\bf 37} 064038 (2010).}

\bibitem{[Bla80]}
{J.P. Blaizot, Phys. Rep. {\bf 64}, 171 (1980).}

\bibitem{[Col04]}
{G. Col\`o, N. and Van Giai, J. Meyer, K. Bennaceur, P. Bonche, Phys. Rev. C
  {\bf 70}, 024307 (2004)}.

\bibitem{[Tod05]}
{B.G. Todd-Rutel and J. Piekarewicz, Phys. Rev. Lett {\bf 95}, 122501 (2005)}.

\bibitem{[Bro98]}
{B.A. Brown, Phys. Rev. {\bf C58}, 220 (1998)}.

\bibitem{[Zuo99]}
{W. Zuo, I. Bombaci, and  U. Lombardo, Phys. Rev. C \textbf{60}, 024605 (1999)}.

\bibitem{[Zuo02]}
{W. Zuo, A. Lejeune, U. Lombardo, and  J.F. Mathiot, Eur. Phys. J. A \textbf{14},
  469 (2002)}.

\bibitem{[van05]}
{E.N.E. van Dalen, C. Fuchs, A. Faessler, Phys. Rev. Lett. {\bf 95}, 022302
  (2005).}

\bibitem{[Heb09]}
{K. Hebeler, T. Duguet, T. Lesinski, and A. Schwenk, Phys. Rev. C {\bf 80},
  044321 (2009).}

\bibitem{[Gor03]}
{S. Goriely, M. Samyn, M. Bender, and J.M. Pearson, Phys. Rev. C {\bf 68},
  054325 (2003)}.

\bibitem{[Li08]}
{B. Li, L. Chen, and C. Ko, Phys. Rep. {\bf 464}, 113 (2009).}

\bibitem{[Ton84]}
{F. Tondeur, M. Brack, M. Farine, and J.M. Pearson, Nucl. Phys. A {\bf 420}, 297
  (1984)}.

\bibitem{[Rei06]}
{P.-G. Reinhard, M. Bender, W. Nazarewicz, and T. Vertse, Phys. Rev. C {\bf
  73}, 014309 (2006).}

\bibitem{[Che05]}
{L. Chen, C. Ko, and B. Li, Phys. Rev. C {\bf 72}, 064309 (2005).}

\bibitem{[Che09]}
{L. Chen, B. Cai, C.M. Ko, B. Li, C. Shen, and J. Xu, Phys. Rev. C {\bf 80},
  014322 (2009).}

\bibitem{[Les07]}
{T. Lesinski, M. Bender, K. Bennaceur, T. Duguet, and J. Meyer, Phys. Rev. C
  {\bf 76}, 014312 (2007)}.

\bibitem{[Dat09]}
{http://orph02.phy.ornl.gov/workshops/lacm08/UNEDF/\-database.html}.

\bibitem{[Aud03]}
{G. Audi, A.H. Wapstra, and C. Thibault, Nucl. Phys. A {\bf 729}, 337 (2003)}.

\bibitem{[Bei75]}
{M. Beiner, H. Flocard, N. Van Giai, and P. Quentin, Nucl. Phys. A {\bf 238},
  29 (1975)}.

\bibitem{[Zal08]}
{M. Zalewski, J. Dobaczewski, W. Satu{\l}a, and T.R. Werner, Phys. Rev. C {\bf
  77}, 024316 (2008)}.

\bibitem{[Aud95]}
{G. Audi and A.H. Wapstra, Nucl. Phys. A {\bf 595}, 409 (1995); Nucl. Phys. A {\bf  565}, 1 (1993)}.

\bibitem{[Fri95]}
{G. Fricke, C. Bernhardt, K. Heilig, L.A. Schaller, L. Schellenberg, E.B.
  Shera, and C.W. De Jager, At. Data Nucl. Data Tables {\bf 60}, 177 (1995)}.

\bibitem{[Ams08]}
{C. Amsler {\it et al.} (Particle Data Group), Phys. Lett. B {\bf 667}, 1
  (2008).}

\bibitem{[Ber09a]}
{G. F. Bertsch, C. A. Bertulani, W. Nazarewicz, N. Schunck and M. V. Stoitsov,
  Phys. Rev. C {\bf 79}, 034306 (2009)}.

\bibitem{[Cao06]}
{L.G. Cao, U. Lombardo, and P. Schuck, Phys. Rev. C {\bf 74}, 064301 (2006).}

\bibitem{[Gor09]}
{S. Goriely, N. Chamel, and J. M. Pearson, Phys. Rev. Lett. {\bf 102}, 152503
  (2009)}.

\bibitem{[Sat98a]}
{W. Satu{\l}a, J. Dobaczewski, and W. Nazarewicz, Phys. Rev. Lett. {\bf 81},
  3599 (1998)}.

\bibitem{[Dob01fw]}
{J. Dobaczewski, P. Magierski, W. Nazarewicz, W. Satu{\l}a, and Z. Szyma\'nski,
  Phys. Rev. C {\bf 63}, 024308 (2001).}

\bibitem{[Dob84]}
{J. Dobaczewski, H. Flocard, and J. Treiner, Nucl. Phys. A {\bf 422}, 103
  (1984)}.

\bibitem{[Dob95c]}
{J. Dobaczewski, W. Nazarewicz, and T.R. Werner, Phys. Scr. {\bf T56}, 15
  (1995)}.

\bibitem{[Kol03]}
{T. Kolda, R.M. Lewis, and V. Torczon, SIAM Review {\bf 45}, 385 (2003).}

\bibitem{(Gol89)}
David~E. Goldberg.
\newblock {\em Genetic Algorithms in Search, Optimization, and Machine
  Learning}.
\newblock Addison-Wesley, Reading, Massachusetts, 1989.

\bibitem{[Mor09]}
{J.J. Mor\'e and S.M. Wild, SIAM J. Optim.{\bf 20}, 172 (2009).}

\bibitem{[Wil08]}
{S.M. Wild, 10th C.M. Conf. Iter. Meth.  (2008).}

\bibitem{(Con09)}
Andrew~R. Conn, Katya Scheinberg, and Lu\'is~N. Vicente.
\newblock {\em Introduction to Derivative-Free Optimization}.
\newblock MPS/SIAM Series on Optimization. Society for Industrial and Applied
  Mathematics, Philadelphia, PA, USA, 2009.

\bibitem{[Sto05]}
{M.V.~Stoitsov, J.~Dobaczewski, W.~Nazarewicz, and P.~Ring, Comput. Phys.
  Commun. {\bf 167}, 43 (2005)}.

\bibitem{[Dob97]}
{J. Dobaczewski and J. Dudek, Comput. Phys. Commun. {\bf 102}, 166 (1997); {\bf
  102}, 183 (1997)}.

\bibitem{[Pei08a]}
{J.C. Pei, M.V. Stoitsov, G.I. Fann, W. Nazarewicz, N. Schunck, and F.R. Xu,
  Phys. Rev. C {\bf 78}, 064306 (2008)}.

\bibitem{[Sto03]}
{M.V. Stoitsov, J. Dobaczewski, W. Nazarewicz, S. Pittel, and D.J. Dean, Phys.
  Rev. C {\bf 68}, 054312 (2003)}.

\bibitem{[Les06]}
{T. Lesinski, K. Bennaceur, T. Duguet and J. Meyer, Phys. Rev. C {\bf 74},
  044315 (2006)}.

\bibitem{[TAO]}
J.~Mor\'e T.~Munson S.~Benson, L. Curfman~McInnes and J.~Sarich.
\newblock Tao user manual (revision 1.9).
\newblock Technical Report ANM/MCS-TM-242, Mathematics and Computer Science
  Division, Argonne National Laboratory, 2007.
\newblock http://www.mcs.anl.gov/tao.

\bibitem{[Ang01a]}
{M. Anguiano, J.L. Egido, and L.M. Robledo, Nucl. Phys. A {\bf 696}, 467
  (2001)}.

\bibitem{[Les08]}
{T. Lesinski, T. Duguet, K. Bennaceur, and J. Meyer, Eur. Phys. J. A {\bf 40},
  121 (2009).}

\bibitem{[Gor06]}
{S. Goriely, M. Samyn, and J.M. Pearson, Nucl. Phys. A {\bf 773}, 279 (2006)}.

\bibitem{[Bla76]}
{J. P. Blaizot, Phys. Lett. B {\bf 60}, 435 (1976)}.

\bibitem{[Cau80a]}
{E. Caurier and B. Grammaticos, Phys. Lett. B {\bf 92}, 236 (1980)}.

\bibitem{[Kor10]}
{M. Kortelainen and T. Lesinski, J. Phys. G: Nucl. Part. Phys. {\bf 37}, 064039
  (2010).}

\bibitem{[Fet71]}
{A.L. Fetter and J.D. Walecka, {\em Quantum Theory of Many-Particle Systems}
  (McGraw-Hill, Boston, 1971)}.

\bibitem{[Gar92b]}
{C. Garc\'{\i}a-Recio, J. Navarri, N. Van Giai, and N. N. Salcedo, Ann. of
  Phys. {\bf 214}, 293 (1992)}.

\bibitem{[Mar06a]}
{J. Margueron, J. Navarro, N. Van Giai, Phys.Rev. C {\bf 74}, 015805 (2006)}.

\bibitem{[Dav09]}
{D. Davesne, M. Martini, K. Bennaceur, J. Meyer, Phys. Rev. C {\bf 80}, 024314
  (2009).}

\bibitem{[Sch07]}
{N. Schwierz, I. Wiedenhover, and A. Volya, arXiv:0709.3525}.

\bibitem{[Woo92]}
{J.L. Wood, K. Heyde, W. Nazarewicz, M. Huyse, and P. van Duppen, Phys. Rep.
  {\bf 215}, 101 (1992)}.

\bibitem{[Rei99]}
{P.-G. Reinhard, D.J. Dean, W. Nazarewicz, J. Dobaczewski, J.A. Maruhn, and
  M.R. Strayer, Phys. Rev. C {\bf 60}, 014316 (1999)}.

\bibitem{[Ska97]}
{J. Skalski, S. Mizutori, and W. Nazarewicz, Nucl. Phys. A {\bf 617}, 282
  (1997)}.
  
\bibitem{[DSN04]} J. Dobaczewski, M.V. Stoitsov, W. Nazarewicz, AIP Conf. Proc. {\bf 726}, 51 (2004); arXiv:nucl-th/0404077v1.


\bibitem{[Naz94a]}
{W. Nazarewicz, Nucl. Phys. A {\bf 574}, 27c (1994)}.

\bibitem{(NLR89)}
George A.~F. Seber, and C.~J. Wild.
\newblock {\em Nonlinear Regression}.
\newblock Wiley, 1989.

\bibitem{[Mon03]}
{D.C. Montgomery and G.C. Taylor, {\it Applied statistics and probability for
  engineers}, John Wiley and Sons (2003).}

\bibitem{[Don87]}
{J.R. Donaldson and R.B. Schnabel, Technometrics {\bf 26}, 67 (1987).}

\bibitem{[Coc04a]}
{B. Cochet, K. Bennaceur, P. Bonche, T. Duguet, and J. Meyer, Nucl. Phys. A
  {\bf 731}, 34 (2004)}.

\bibitem{[War09]}
{M. Warda, X. Vi\~nas, X. Roca-Maza, and M. Centelles, Phys. Rev. C {\bf 80},
  024316 (2009).}

\end{thebibliography}

\end{document}